\documentclass[]{aastex63}

\usepackage{gensymb}

\def\Msun{\mbox{${\rm M}_{\odot}$}}
\def\Rsun{\mbox{${\rm R}_{\odot}$}}
\def\Lsun{\mbox{${\rm L}_{\odot}$}}
\def\kms{\mbox{km\,s$^{-1}$}}
\usepackage{xspace}

\received{May 3, 2021}
\revised{May 25, 2021}
\accepted{May 26, 2021}
\submitjournal{ApJ}

\shorttitle{$\nu$~Gem: a triple system with an outer Be star}
\shortauthors{Klement et al.}

\graphicspath{{./}{figures/}}

\begin{document}

\title{$\nu$ Gem: a hierarchical triple system with an outer Be star}

\correspondingauthor{Robert Klement}
\email{robertklement@gmail.com}

\author[0000-0002-4313-0169]{Robert Klement}
\affiliation{The CHARA Array of Georgia State University, Mount Wilson Observatory, Mount Wilson, CA 91023, USA}
\author[0000-0002-4518-3918]{Petr Hadrava}
\affiliation{Astronomical Institute, Academy of Sciences of the Czech Republic, Bo\v{c}n\'{i} II 1401, CZ 141 31 Praha 4, Czech Republic}
\author[0000-0003-1013-5243]{Thomas Rivinius}
\affiliation{European Organisation for Astronomical Research in the Southern Hemisphere (ESO), Casilla 19001, Santiago 19, Chile}
\author[0000-0003-1637-9679]{Dietrich Baade}
\affiliation{European Organisation for Astronomical Research in the Southern Hemisphere (ESO), \\ Karl-Schwarzschild-Str.\ 2, 85748 Garching bei M\"unchen, Germany}
\author[0000-0003-2050-1227]{Mauricio Cabezas}
\affiliation{Astronomical Institute, Academy of Sciences of the Czech Republic, Bo\v{c}n\'{i} II 1401, CZ 141 31 Praha 4, Czech Republic}
\affiliation{Institute of Theoretical Physics, Faculty of Mathematics and Physics, Charles University, V Holesovickach 2 \\
Prague, 180 00, Czech Republic}
\author[0000-0002-1082-7496]{Marianne Heida}
\affiliation{European Organisation for Astronomical Research in the Southern Hemisphere (ESO), \\ Karl-Schwarzschild-Str.\ 2, 85748 Garching bei M\"unchen, Germany}
\author[0000-0001-5415-9189]{Gail H. Schaefer}
\affiliation{The CHARA Array of Georgia State University, Mount Wilson Observatory, Mount Wilson, CA 91023, USA}
\author[0000-0002-3003-3183]{Tyler Gardner}
\affiliation{Department of Astronomy, University of Michigan, 1085 S. University Ave, Ann Arbor, MI 48109, USA}
\author[0000-0001-8537-3583]{Douglas R. Gies}
\affiliation{Center for High Angular Resolution Astronomy and
  Department of Physics and Astronomy,\\
  Georgia State University, P. O. Box 5060, Atlanta, GA 30302-5060, USA}
\author[0000-0002-2208-6541]{Narsireddy Anugu}
\affiliation{Steward Observatory, Department of Astronomy, University of Arizona, 933 N. Cherry Ave, Tucson, AZ 85721, USA}
\author[0000-0001-9745-5834]{Cyprien Lanthermann}
\affiliation{The CHARA Array of Georgia State University, Mount Wilson Observatory, Mount Wilson, CA 91023, USA}
\author[0000-0001-9764-2357]{Claire L. Davies}
\affiliation{Astrophysics Group, School of Physics, University of Exeter, Stocker Road, Exeter, EX4 4QL, UK}
\author[0000-0002-9759-038X]{Matthew D. Anderson}
\affiliation{The CHARA Array of Georgia State University, Mount Wilson Observatory, Mount Wilson, CA 91023, USA}
\author[0000-0002-3380-3307]{John D. Monnier}
\affiliation{Department of Astronomy, University of Michigan, 1085 S. University Ave, Ann Arbor, MI 48109, USA}
\author[0000-0002-1575-4310]{Jacob Ennis}
\affiliation{Department of Astronomy, University of Michigan, 1085 S. University Ave, Ann Arbor, MI 48109, USA}
\author[0000-0001-8837-7045]{Aaron Labdon}
\affiliation{Astrophysics Group, School of Physics, University of Exeter, Stocker Road, Exeter, EX4 4QL, UK}

\author[0000-0001-5980-0246]{Benjamin R. Setterholm}
\affiliation{Department of Astronomy, University of Michigan, 1085 S. University Ave, Ann Arbor, MI 48109, USA}
\author[0000-0001-6017-8773]{Stefan Kraus}
\affiliation{Astrophysics Group, School of Physics, University of Exeter, Stocker Road, Exeter, EX4 4QL, UK}
\author[0000-0002-0114-7915]{Theo A. ten Brummelaar}
\affiliation{The CHARA Array of Georgia State University, Mount Wilson Observatory, Mount Wilson, CA 91023, USA}
\author[0000-0002-0493-4674]{Jean-Baptiste Le Bouquin}
\affiliation{Department of Astronomy, University of Michigan, 1085 S. University Ave, Ann Arbor, MI 48109, USA}
\affiliation{Institut de Planétologie et d'Astrophysique de Grenoble, CNRS, Univ. Grenoble Alpes, France}

\begin{abstract}
Time series of spectroscopic, speckle-interferometric, and optical long-baseline-interferometric observations confirm that $\nu$~Gem is a hierarchical triple system.  It consists of an inner binary composed of two B-type stars and an outer classical Be star. Several photospheric spectral lines of the inner components were disentangled, revealing two stars with very different rotational broadening ($\sim$260 and $\sim$140\,\kms, respectively), while the photospheric lines of the Be star remain undetected. From the combined spectroscopic and astrometric orbital solution it is not possible to unambiguously cross-identify the inner astrometric components with the spectroscopic components. In the preferred solution based on modeling of the disentangled line profiles, the inner binary is composed of two stars with nearly identical masses of 3.3\,M$_\odot$ and the more rapidly rotating star is the fainter one. These two stars are in a marginally elliptical orbit ($e$ = 0.06) about each other with a period of 53.8\,d.  The third star also has a mass of 3.3\,\Msun\ and follows a more eccentric ($e$ = 0.24) orbit with a period of 19.1\,yr.  The two orbits are co-directional and, at inclinations of 79\degree\ and 76\,\degree\ of the inner and the outer orbit, respectively, about coplanar.  No astrometric or spectroscopic evidence could be found that the Be star itself is double.   The system appears dynamically stable and not subject to eccentric Lidov-Kozai oscillations.  After disentangling, the spectra of the components of the inner binary do not exhibit peculiarities that would be indicative of past interactions.  Motivations for a wide range of follow-up studies are suggested.  
\end{abstract}

\keywords{Be stars (142) --  Circumstellar disks (235) -- Optical interferometry (1168) -- Orbit determination (1175) -- Spectroscopy (1558) -- Trinary stars (1714)}

\section{Introduction} 
\label{sec:intro}

Be stars occupy the main sequence (MS) in its full width and from late O to early A stars.  This part of the Hertzsprung-Russell (HR) diagram is home to many physically important phenomena.  Stars at the hot and massive end will evolve to core-collapse supernovae, and this domain is intersected by the $\beta$\,Cephei instability strip \citep{1992A&A...256L...5M}.  At the cool end, several classes of chemically peculiar stars are found.  Slowly Pulsating B (SPB) Stars populate the mid range from B2 to B9 \citep{2007MNRAS.375L..21M}, and the fraction of stars in binaries increases from A to O stars across the B-star domain \citep{2017ApJS..230...15M}.  About 7\% of all Galactic OB stars have major magnetic dipole fields \citep{2014IAUS..302..265W}.  

Be stars share many of these properties but not all.  Most notably, there are no Be stars with detected magnetic fields \citep{2016ASPC..506..207W}, and the fraction of Be stars with main-sequence (MS) companions seems vanishingly low at least for early-type Be stars \citep{2020A&A...641A..42B}. The most conspicuous difference between Be and other B stars is that Be stars are surrounded by a Keplerian gaseous disk, where the eponymous emission lines form \citep{2013A&ARv..21...69R}.   On average, Be stars rotate more rapidly than most other B stars which is generally understood as a necessary condition for the formation of the disk.  The star-to-disk mass transfer appears dominated by discrete outbursts which are more prevalent among early-type Be stars \citep{LabadieB2017, 2018MNRAS.479.2909B}.  The nonlinear superposition of multiple nonradial pulsation modes found in long-term space photometry \citep{2018pas8.conf...69B} is the strongest observed contender for powering the outbursts, although magnetic fields not detectable with current means have also been postulated \citep{2020MNRAS.493.2528B}.  

The lack of MS companions and the elevated rotation rates may be related in that Be stars may have been spun up by mass and angular-momentum transfer from an initially more massive companion that has evolved to a hot low-mass star \citep{1991A&A...241..419P, 2014ApJ...796...37S}, as is the case in Be X-ray binaries in which the secondary stars are more massive neutron-stars \citep{2011Ap&SS.332....1R}. To shed further light on the role of binarity in the genesis of Be stars, it is useful to study stellar systems consisting of more than two stars.  Only a dozen of such systems have been identified to date \citep[][Rivinius et al., in prep]{1984ApJ...285..190A}.  At $V$=4.14\,mag and $H$=4.43\,mag, \object{$\nu$\,Gem} is the apparently brightest of them.  

This paper presents $\nu$\,Gem's  hierarchical-triple system architecture, the orbital parameters, and the properties of its components.   After Sect.\,\ref{sec:literature} has introduced $\nu$~Gem from the literature, Sect.\,\ref{sec:observations} describes the additional observations used in this study.  Sect.\,\ref{sec:methodology} explains the methods applied for the extraction of spectral and orbital information from the observations, and the analysis of the results is performed in Sect.\,\ref{sec:analysis}.  The discussion and the conclusions follow in Sects.\,\ref{sec:discussion} and \ref{sec:conclusions}, respectively.

\section{$\nu$~Gem in the literature}
\label{sec:literature}

The first indication that \object{$\nu$\,Gem} (HD\,45542, HR\,2343, HIP\,30883, WDSJ06290+2013) is not single was reported by \citet{HarperNuGem} who derived a spectroscopic period of 9.6 years, which is close to one-half of the true period of 19.14 years determined below.  He noted `a complex character of the lines'. However, H$\alpha$ was not included in the spectral range, and \citeauthor{HarperNuGem} did not mention line emission. He added `that there is a short-period oscillation in the velocities appears without doubt' but called the velocity range of the latter `small'.  Prompted by the reported $a \sin i$ value of $1.417 \times 10^9$\,km, \citet{1920AN....211...13B} tried to resolve the presumed pair visually, also using an absorption wedge, and found the images to be elongated.  Their estimate of the separation of $\sim0.15$\,arcsec is larger than, but not demonstrably grossly discrepant with, the maximum angular distance between components A and B measured by speckle interferometry decades later.  Sect.\,\ref{sec:speckle} analyzes the long series of speckle observations.  Only the optical long-baseline interferometry (OLBI) presented in Sect.\,\ref{sec:OLBI} could resolve the brighter component into two stars. 

For a long time, the composite nature of the spectra caused confusion to observers who were not aware of it.  For instance, \citet{1982ApJS...50...55S} listed $\nu$~Gem with a $v$\,sin\,$i$ of 170\,km\,s$^{-1}$ whereas the profiles published by \citet{1996AAS..116..309H} show unmistakable Be-shell characteristics.  \citet{1973ApJ...179..221S} even listed $\nu$~Gem as `true pole-on star'. These results are incompatible for single Be stars because shell stars, in which the line of sight intersects the equatorial circumstellar disk, are viewed equator-on whereas 170\,km\,s$^{-1}$ is only about one-half of the equatorial rotational velocity of mid- to late-type Be stars.  \citet{2006A&A...459..137R} resolved this conflict with a time series of optical spectra:  In addition to the shell absorption in H$\alpha$, there are absorption lines that are not associated with the Be star:  They are due to a second star and shift with a period of 53.72\,d.  However, the Be star does not exhibit orbital reflex motion with the same period.  On this timescale, the Be star's radial velocity (RV) is roughly constant.  Accordingly, $\nu$\,Gem could be a triple system in which the outer Be star orbits an inner binary.  For lack of spectral lines detected from the third star, the physical association of the Be star with the binary remained unclear.  The final confirmation came only recently from long-baseline optical interferometry \citep{2021AJ....161...40G} which also resolved inconsistencies in previous determinations from speckle observations of the orbital period of the Be star.  

Unlike some other Be stars, $\nu$~Gem was apparently not observed (at least since 1915) in a state free of H$\alpha$ line emission \citep[cf.][and references therein]{1973ApJ...179..221S, 1982ApJS...50...55S}.  That is, the Be star always possessed a well-developed circumstellar disk.  Without complex modeling, the variability of the H$\alpha$ line profile of $\nu$~Gem is easily misinterpreted.  For instance, the central depression is sometimes considerably deeper than the ambient continuum \citep{1996AAS..116..309H, 2006A&A...459..137R, 2013A&A...550A..79C} and satisfies even the strict definition by \citet{1996A&A...308..170H} of Be-shell stars.  On the one hand, this depth is surprising because shell features should be diluted in the combined light of three stars.  Dilution probably contributed to the finding by \citet{1996AAS..116..309H} that, among all Be-shell stars observed by them, $\nu$~Gem had the weakest H$\alpha$ line emission. \citet{2006A&A...459..137R} observed indications of dilution in very narrow \ion{Fe}{2} lines.  These line profiles demonstrate the shell-star nature beyond the confusion potential of the H$\alpha$ variability although at times these lines were washed out, presumably by disk oscillations or by blending with photospheric lines from the inner binary or both.  On the other hand, a narrow absorption core from another star can deepen the central reversal in H$\alpha$.  To make things even more complicated, \citeauthor{2006A&A...459..137R} also recorded a transient flat-top structure in H$\alpha$ with two absorption and three emission wiggles, without any obvious trace of a shell absorption.  Such emission profiles can arise in Be disks truncated by a companion star and viewed at angles around 60$^{\rm o}$ \citep{2018MNRAS.473.3039P}, for which disk models do not predict strong shell absorptions.  This structure persisted for two months so that its physical counterpart must be localized outside the inner binary which has an orbital period of 53.7\,d.  

At the same time when the emission in H$\alpha$ exhibited a flat-topped profile without indication of a superimposed shell absorption component, the shell character of the \ion{Fe}{2}\,$\lambda 5169$ line was no longer detectable either.  $\nu$~Gem might, therefore, be one of the rare Be stars that show transitions between Be and Be-shell phases \citep{2006A&A...459..137R} as proposed by \citet{2014ApJ...795...82S}.  These transitions are unexplained and may involve geometric thickness variations or precession of the disk in a binary \citep{2007ASPC..361..267H}. \citet{2014ApJ...795...82S} determined an inclination angle of 77$^\circ$ which is still compatible with the typical disk opening angle of $2 \times 13^\circ$ (measured from midplane) estimated by \citet{1996A&A...308..170H}.  Accordingly, geometric thickness variations are a plausible cause of the transient shell characteristics of $\nu$~Gem. 

The H$\alpha$ emission profiles also exhibited pronounced symmetry variations in the violet-to-red intensity ratio of their double peaks \citep{1996A&AS..116..309H, 2006A&A...459..137R} which are commonplace in Be stars \citep{1997A&A...318..548O}.  The timescales can be years if these so-called $V/R$ variations are due to one-armed disk oscillations driven by the rotationally induced gravitational quadrupole moment of the Be star.  They can also correspond to half the orbital period of a companion star if two-armed oscillations are resonantly excited by the latter. Not seldom, resonant excitation of disk oscillations and disk truncation occur together \citep{2018MNRAS.473.3039P}.  If such processes affect the disk of the Be star in $\nu$~Gem, it is not clear whether the rather distant inner binary can be responsible for them.  \citet{2018ApJ...853..156W} searched for ultraviolet (UV) signatures of a hot, subluminous companion of the Be star.  However, the quality of the available spectra was insufficient for a meaningful analysis.  $\nu$~Gem was not included in the study by \citet{Klement2019} who investigated the spectral energy distribution of Be stars for turndowns as tracers of disk truncation by a companion.  

Whereas \citet{2006A&A...459..137R} did not observe line emission in \ion{Fe}{2}\,$\lambda 5169$, \citet{1996AAS..116..309H} presented a profile with a single red peak at $\sim$3\% of the continuum flux. The red edge of this feature was extremely sharp with a half width probably not exceeding a few km\,s$^{-1}$.  Although similar profiles exist for quite a few other Be stars in the atlas of \citeauthor{1996AAS..116..309H}, they are unexplained.  As in some of the spectra observed by \citeauthor{2006A&A...459..137R}, the profile did not include an obvious shell component, but there was a much broader blue-shifted absorption feature.  

For the understanding of the temporary appearance of shell absorption lines in $\nu$~Gem a comparison to other $V/R$-variable Be stars is useful.  Very similar variability of circumstellar lines is exhibited by the B1 shell star \object{$\zeta$\,Tau}.  This star is a 133-day binary with an unseen companion and at times undergoes $V/R$ variations with cycle lengths of a few years \citep{2009A&A...504..929S}.  Its inclination angle has been determined as 85$^\circ$ \citep{2009A&A...504..915C}.  In $\zeta$\,Tau, the presence of flat-topped H$\alpha$ emission and both narrow and broad absorption profiles of typical shell lines is coupled to specific phases of the $V/R$ cycles \citep{2009A&A...504..929S}. $V/R$ variability \citep[see][]{2002ASPC..279..221H} as well as flat-topped H$\alpha$ profiles \citep{Borre2020} have been observed also in the B0.5e star \object{$\gamma$\,Cas}, similarly to both $\nu$~Gem and $\zeta$~Tau.  $\gamma$\,Cas also has an unseen companion \citep{2012A&A...537A..59N}, but is viewed at an inclination angle of $\sim$42$^\circ$ \citep{2012A&A...545A..59S} so that shell lines are not normally expected. Nevertheless,  \citet{1994A&A...284..515T} found a correlation between the $V/R$ variability in H$\beta$ with the strength of discrete absorption components of stellar wind lines which also varied with $V/R$ phase.  \citeauthor{1994A&A...284..515T} concluded that the one-armed disk oscillations that give rise to the $V/R$ variability also modulate the geometric thickness of the disk.  As the result, the wind appears stronger when the thicker parts of the disk reach closer to the line of sight and radiative ablation of the disk becomes more visible in the profiles of UV wind lines.  This notion  also explains the cyclic appearance of shell features in $\zeta$\,Tau and is probably applicable to $\nu$~Gem, too.  In any event, the close qualitative agreement of the spectral variations seen in $\nu$~Gem and $\zeta$\,Tau suggests that those of $\nu$~Gem are not drastically contaminated by the other two stars in the system.  An intriguing but unanswered question is whether flat-topped H$\alpha$ emission profiles are unequivocal indicators of binarity.  If yes, and the distant inner binary in $\nu$~Gem is not the reason, the Be star would itself be double.  

The continuum flux from normal stars is usually unpolarized, whereas this is different for Be stars, mostly due to scattering in the point-symmetry-breaking disk \citep{1976ApJ...206..182P, 2013ApJ...765...17H, 2013A&ARv..21...69R}. NASA's Astrophysics Data System \citep[ADS;][]{2000A&AS..143...41K} lists six papers with broad-band linear polarimetry of $\nu$~Gem.  The degree of the polarization was a few 0.1\%. Small variability demonstrates that the polarization is not entirely due to interstellar dust but intrinsic to at least one component of the system.  Most probably, the polarization arises from the circumstellar disk of the Be star, but an additional asymmetry, and therefore source of polarization, can be the inner binary.  Accordingly, only changes in the polarization angle (if present) can be interpreted without tremendous observational effort.  The polarization angles range from 11 to 37 degrees.  In view of the reported uncertainties of up to 5 degrees, this spread probably does not indicate a change in orientation of the disk.  While the polarization degree was independent of wavelength between 400 and 950\,nm in the data of \citet{1979AJ.....84..812P}, \citet{1999AJ....118.1061G} found a decrease with increasing wavelength.

\section{Observations}
\label{sec:observations}

\subsection{Astrometry}

In the astrometric notation adopted for this study, the brighter inner component is referred to as Aa, the fainter inner component as Ab, and the component on the wide orbit as B. 

\subsubsection{Speckle interferometry}
\label{sec:speckle}

The wide orbit of the outer Be star (B) around the center of light of the inner binary (Aa+Ab) was constrained by several speckle observing campaigns, and the most recent set of astrometric orbital elements was published by \citet{2008NewA...13..587C}. The Sixth Catalog of Orbits of Visual Binary Stars\footnote{\url{http://www.astro.gsu.edu/wds/orb6.html}} assigns this orbital solution for $\nu$~Gem grade 2, meaning that no major changes in the orbital elements are likely to occur when new data become available. We recovered a set of 31 speckle interferometry measurements of the component separation and position angle from the Washington Double Star Catalog\footnote{\url{http://www.astro.gsu.edu/wds/}} \citep[WDS,][]{2019yCat....102026M}, covering the years 1976 to 2016, and including 11 measurements taken after the determination of the orbital solution by \citet{2008NewA...13..587C}. The previously published speckle measurements corrected for the precession of North are listed in Table~\ref{tab:speckle}. The typical 1-$\sigma$ uncertainties of the speckle measurements are of the order of several milliarcseconds (mas), when available.

\subsubsection{Optical long baseline interferometry}
\label{sec:OLBI}

The Center for High Angular Resolution Astronomy (CHARA) Array is a 6-telescope optical/near-IR interferometer with a maximum baseline of $B_\mathrm{max}\sim330$\,m \citep{2005ApJ...628..453T}. The MIRC-X beam combiner \citep{2020AJ....160..158A} - since 2017 September an upgraded version of the former MIRC beam combiner \citep{2006SPIE.6268E..1PM} - is capable of combining coherently all six beams in the $H$-band ($\lambda = 1.6\,\mu$m), achieving a good $(u,v)$ plane coverage, even with snapshot observations, and an angular resolution $\lambda / (2 B_\mathrm{max})$ of 0.5\,mas. Both MIRC/MIRC-X are capable of astrometry precision of $<10$\,micro-arcsec \citep[$\mu$as;][]{2018ApJ...855....1G, 2021AJ....161...40G}.

In this study we make use of two archival measurements of $\nu$~Gem taken by MIRC in 2015/2016, and eight measurements by MIRC-X obtained via the ARrangement for Micro-Arcsecond Differential Astrometry (ARMADA) survey \citep{2021AJ....161...40G}. For extensive details on the MIRC/MIRC-X observations, data reduction, and extraction of the astrometric positions from the data, we refer the reader to the dedicated ARMADA paper \citep{2021AJ....161...40G}. 

All three point source components were detected on both occasions by MIRC, and on two occasions by MIRC-X. For the remaining six MIRC-X measurements, the fainter component Ab in the inner binary was not reliably detected due to poor calibration and group delay tracking, and only the separation of the outer pair Aa+B could be determined. The measurements are listed in Table~\ref{tab:astro-mirc-inner} for the inner binary Aa+Ab and in Table~\ref{tab:astro-mirc-outer} for the outer Aa+B pair.

\begin{deluxetable}{lLCCCCC}
\tablecaption{MIRC (first two epochs) and MIRC-X (last two epochs) astrometry of the inner pair Aa+Ab. \label{tab:astro-mirc-inner}}
\tablewidth{0pt}
\tablehead{
\colhead{UT Date} & \colhead{MJD} & \colhead{separation} & \colhead{PA} & \colhead{$\sigma$-major$^\mathrm{a}$} & \colhead{$\sigma$-minor$^\mathrm{a}$} & \colhead{$\sigma$-PA$^\mathrm{a}$}\\
\nocolhead{Date} & \nocolhead{MJD} & \colhead{[mas]} & \colhead{[\degree]} & \colhead{[mas]} & \colhead{[mas]} & \colhead{[\degree]} 
}
\startdata
2015Nov23 & 57349.375 & 1.850 & 118.823 & 0.010 & 0.008 & 118.5216\\
2016Nov14 & 57706.434 & 2.965 & 312.417 & 0.012 & 0.008 & 141.4462\\
2017Sep28 & 58024.562 & 2.777 & 307.268 & 0.055 & 0.028 & 286.86\\
2017Sep30 & 58026.549 & 2.923 & 309.670 & 0.043 & 0.023 & 305.08\\
\enddata
\tablecomments{a - Parameters of the error ellipses with $\sigma$-PA being the PA of the major axis of the error ellipse (measured from North to East).}
\end{deluxetable}

\begin{deluxetable}{lCCCCCC}
\tablecaption{MIRC (first two epochs) and MIRC-X astrometry of the outer pair Aa+B. \label{tab:astro-mirc-outer}}
\tablewidth{0pt}
\tablehead{
\colhead{UT Date} & \colhead{MJD} & \colhead{separation} & \colhead{PA} & \colhead{$\sigma$-major$^\mathrm{a}$} & \colhead{$\sigma$-minor$^\mathrm{a}$} & \colhead{$\sigma$-PA$^\mathrm{a}$}\\
\nocolhead{Date} & \nocolhead{MJD} & \colhead{[mas]} & \colhead{[\degree]} & \colhead{[mas]} & \colhead{[mas]} & \colhead{[\degree]} 
}
\startdata
2015Nov23 & 57349.375 & 77.809 & 114.589 & 0.400$^\mathrm{b}$ & 0.400$^\mathrm{b}$ & 118.3\\
2016Nov14 & 57706.434 & 87.121 & 118.536 & 0.400$^\mathrm{b}$ & 0.400$^\mathrm{b}$ & 109.6\\
2017Sep28 & 58024.561 & 93.148 & 121.903 & 0.053 & 0.029 & 90.00\\
2017Sep30 & 58026.549 & 93.201 & 121.905 & 0.041 & 0.018 & 121.67\\
2018Sep20 & 58381.505 & 95.478 & 125.440 & 0.159 & 0.029 & 89.93\\
2018Nov21 & 58443.462 & 93.947 & 125.936 & 0.093 & 0.032 & 42.83\\
2018Dec04 & 58456.394 & 92.303 & 125.955 & 0.114 & 0.049 & 137.61\\
2019Sep08 & 58734.537 & 88.279 & 128.610 & 0.059 & 0.034 & 348.69\\
2019Oct13 & 58769.510 & 87.900 & 129.244 & 0.028 & 0.026 & 25.13\\
2019Nov11 & 58798.476 & 88.462 & 129.284 & 0.047 & 0.040 & 47.34\\
\enddata
\tablecomments{a - Parameters of the error ellipses with $\sigma$-PA being the PA of the major axis of the error ellipse (measured from North to East). b - Errors were set to reflect the largest possible wavelength scale correction for data taken without wavelength calibration using a dedicated etalon.}
\end{deluxetable}

The MIRC/MIRC-X measurements in which all three components were detected provide estimates of the relative $H$-band fluxes of the three components. The ratios listed in Table~\ref{tab:rel_flux} were derived from the best-calibrated observation, namely the MIRC dataset taken on 2015-Nov-23. The values are in agreement with the magnitude difference between the inner pair and the outer star reported by \citet{2020AJ....159..233H}.

\begin{deluxetable*}{CCCCC}
\tablecaption{Relative $H$-band fluxes of the three components extracted from MIRC interferometry. \label{tab:rel_flux}}
\tablewidth{0pt}
\tablehead{
\colhead{$F_\mathrm{Aa}$} & \colhead{$F_\mathrm{Ab}$} & \colhead{$F_\mathrm{B}$} & \colhead{$F_\mathrm{Aa} / F_\mathrm{Ab}$} & \colhead{$F_\mathrm{Aa} / F_\mathrm{B}$}
}
\startdata
0.437\pm0.009 & 0.224\pm0.004 & 0.339\pm0.008 & 1.95\pm0.05 & 1.29\pm0.04\\
\enddata
\end{deluxetable*}

\subsection{Spectroscopy}

The main spectroscopic datasets consist of spectra from the coud\'e spectrograph of the Ond\v rejov 2-m telescope\footnote{\url{https://stelweb.asu.cas.cz/web/index.php?pg=2m_telescope}} covering a region around the H$\alpha$ line, and from the {\sc Heros} echelle spectrograph\footnote{\url{https://www.lsw.uni-heidelberg.de/projects/instrumentation/Heros/}}. A more detailed description of the {\sc Heros} instrument and the reduction procedure can be found in \citet{2001JAD.....7....5R} and references therein. We also make use of eight exposures with the ESPaDOns spectrograph\footnote{\url{http://www.ast.obs-mip.fr/projets/espadons/espadons.html}} taken in a single observing night. In addition, we retrieved spectra from the BeSS database\footnote{\url{http://basebe.obspm.fr/basebe/}} of spectra taken predominantly by amateur observers. The BeSS data are of varying quality and resolution and mostly cover the H$\alpha$ region only. 

The spectroscopy analyzed in this study is summarized  in Table~\ref{tab:spectroscopy}. The portions of the observed spectra that were eventually selected for the analysis were normalized using spline fitting of the surrounding continuum. A subset of 14 Ond\v rejov coud\'e and 35 {\sc Heros} spectra was previously analyzed by \citet{2006A&A...459..137R}, who derived an SB1 orbital solution for the inner binary from the \ion{He}{1}\,$\lambda 6678$ and \ion{Si}{2}\,$\lambda 6347$ lines. 

\begin{deluxetable*}{lccCCc}
\tablecaption{Spectroscopic datasets\label{tab:spectroscopy}}
\tablewidth{0pt}
\tablehead{
\colhead{Instrument/Detector} & \colhead{Telescope} & \colhead{MJD} & \colhead{Number} & \colhead{Resolving power} & \colhead{Spectral region}\\
\nocolhead{Instrument} & \nocolhead{Telescope} & \nocolhead{MJD} & \colhead{of spectra} & \colhead{$\lambda / \Delta \lambda$} & \colhead{[$\AA$]} 
}
\startdata
{\sc Heros} blue & Multiple$^\mathrm{a}$ & 51046--52724 & 35 & 20000 & 3500--5750  \\
{\sc Heros} red & Multiple$^\mathrm{a}$ & 51046--52724 & 34 & 20000 & 5750--8600  \\
Coud\'{e} spectrograph / Reticon & Ond\v{r}ejov 2-m & 49658--51464 & 14 & 8000 & 6300--6740  \\
Coud\'{e} spectrograph / CCD & Ond\v{r}ejov 2-m & 52228--55470 & 89 & 13000 & 6300--6740  \\
ESPaDOns & CFHT & 55222 & 8 & 68000 & 3700--8870 \\
BeSS database & Multiple & 54913--58842 & 106 & 5000-30000 & H$\alpha$ \\
\enddata
\tablecomments{a - Calar Alto (1998); Wendelstein (2000); Ondřejov (2000--2003)}
\end{deluxetable*}

We also utilize five UV spectra from the International Ultraviolet Explorer (IUE) that are available at the INES Archive Data Server\footnote{\url{http://sdc.cab.inta-csic.es/ines/index2.html}}.  They were taken through a large aperture and therefore have reliable flux calibration.

\subsection{Polarimetry}

Archival polarization data are available from the HPOL spectropolarimeter \citep{2014JAI.....350009D} and were collected at Pine Bluff Observatory. The actual data can be obtained from the MAST archive, but for the purpose of this work synthesized broad-band filter data are sufficient, which are available at {\tt http://www.sal.wisc.edu/HPOL/tgts/Nu-Gem.html} and listed in Table~\ref{tab:hpol}. 

The star was observed by HPOL at nine epochs, from 1992 to 2004. Data are given for the \textit{UxBVRI}-bands, but the values for {\it Ux} and partly $I$ have a considerably larger scatter than those for {\it BVR}. In the entire 12 years of observation, the polarization degree in these three bands was between 0.25 and 0.3\%, and the position angle was between $15^\circ$ and $20^\circ$. This is in good agreement with the literature values (cf.\ Sect.\,\ref{sec:literature}).  There is no variability beyond the observational errors, confirming the conclusion that the orientation of the disk around the Be stars in $\nu$~Gem has been stable.  This is also in agreement with observational characteristics of $\zeta$\,Tau which at times exhibits very similar variability of its H$\alpha$ profile (cf.\,Sect.~\ref{sec:literature}) and, like $\nu$~Gem, was observed to maintain a constant polarization percentage and angle even through large-amplitude $V/R$ cycles \citep{2009A&A...504..929S, 2009A&A...504..915C}.

Archival ESPaDOnS data (see below) were also taken for the purpose of linear polarimetry, but ESPaDOnS is not capable of measuring absolute polarization values. A very small spectropolarimetric signature relative to the continuum can be detected across the H$\alpha$ line.  However, it is too weak, and yet too structured, to be interpreted straightforwardly (see Fig.~\ref{fig:ESPpol}), and therefore not further considered in the present work.

\subsection{Space photometry}

$\nu$\,Gem is not known as photometrically variable. Archival photometric data, suitable for a period search at the millimagnitude level, exist from the Solar Mass Ejection Imager (SMEI) mission \citep{2013SSRv..180....1H} and were analyzed to check the photometric stability. In these data, one coherent frequency is actually present, $f=2.4068$ cycles per day, with an amplitude of about 2\,mmag. This is not unusual for a mid-type B star, indeed there might be many more, lower-amplitude frequencies as implied by the large sample of Be-star light curves collected by the Terrestrial Exoplanet Survey Satellite  \citep[TESS;][]{2020arXiv201013905L}. Unfortunately, in the first scan by TESS of the northern ecliptic hemisphere, $\nu$~Gem fell into a gap of the raster.  The variability detected with SMEI might originate in any of the three components of the system, and only long-term Doppler frequency shifts may lift this indeterminacy. This has no impact on the following analysis so that it is ignored.

\section{Methods used for data analysis}
\label{sec:methodology}

\subsection{Disentangling of spectra}

Disentangling of spectra is a technique developed for separating component spectra of multiple stellar systems, while simultaneously solving for the orbital elements even without prior knowledge about the orbit or the radial velocities (RVs) of the components \citep{1994A&A...281..286S, 1995A&AS..114..393H}. A measurement of the RVs is an optional additional step in this method. It is performed by fitting each spectrum as the sum of the previously disentangled line profiles and optimizing their Doppler-shifts independently of the orbital solution. The standard disentangling method only finds relative Doppler shifts of the spectra of components in different phases without any need of a template or an identification of the spectral lines. It thus cannot provide the systemic ($\gamma$-) velocity without using synthetic templates for the component spectra. Since the disentangling is performed on normalized spectra, it also does not constrain the component flux ratio, and the disentangled line profiles are normalized to the composite continuum.

To the spectra of $\nu$~Gem, we applied the method of disentangling in the Fourier domain as implemented in the {\sc Korel} code, which supports the analysis of stellar systems with up to five components in a hierarchical architecture \citep{1995A&AS..114..393H, 2004PAICz..92...15H}. This step requires the input spectra to be rebinned to a logarithmic wavelength scale (with identical start values) so that the  RV shift per bin is constant. In our application to $\nu$~Gem, we use {\sc Korel} to (1) disentangle the component line profiles so that stellar properties can be constrained, (2) to determine spectroscopic orbital parameters, and (3) to measure the RVs of the disentangled components, so that they can be combined with the astrometric data for a full orbital solution. The errors of individual parameters are calculated as the maximum 1-$\sigma$ Bayesian probabilities of the solution in two-dimensional cross-sections of the parameter space \citep[with free line-strength factors, cf.][]{2016ASSL..439..113H}. In Sect.~\ref{sec:disentangling_results}, we present the results from {\sc Korel} disentangling of three spectral regions.

\subsection{Simultaneous solution of RV curves and visual orbits}

The orbital motions in the $\nu$~Gem system were represented by two Keplerian orbits, while ignoring the possible dynamical interaction between the orbits over the time spanned by the data. To obtain the orbital parameters of the inner and outer orbits, we used extended versions (cf.\ below) of the {\sc Fotel} code \citep{2004PAICz..92....1H}.

{\sc Fotel} determines an orbit by minimizing $\Sigma{(O-C)^2}$, i.e., the sum of the squares of the residuals (observed minus calculated), as a function of the orbital parameters. The depth and shape of the minimum are then used to estimate the 1-$\sigma$ uncertainties of the fitted parameters. To put the residuals of different observables on a roughly common scale, each quantity is transformed to characteristic units.  The RVs are scaled to the semi-amplitude of the RV-curve  (which is nearly the same for both stars in the inner binary), the position angles (PAs) are expressed in radians (i.e., in units of the order of the whole circle) and the angular distances of the components enter on a logarithmic scale (similarly to the magnitudes used for light curves, the solution of which is the original purpose of {\sc Fotel}). 

Since different orbital parameters are sensitive to different types of input data, it is necessary to combine different datasets simultaneously in order to obtain a full and correct orbital description. In our case, the RVs provide unique information about the projection of the orbit onto the line of sight, and they are the exclusive means to discriminate the semi-amplitudes $K$ of the components' RV-curves and the systemic radial velocity $\gamma$. The astrometry gives the projection of the orbit onto the plane of the sky that is perpendicular to the line of sight.  It is the only source of information about the inclination of the orbit $i$, the angular size $\bar{a}$ of the orbit and the position angle (PA) of the ascending node $\Omega$ (however, the RVs are needed to distinguish the latter from the descending node). RV as well as astrometric data can independently determine the remaining orbital parameters -- the period $P$, the epoch of periastron passage $T$, the eccentricity $e$, and the longitude of periastron $\omega$. This partial redundancy enables us to estimate to what extent the RVs and the astrometry contradict or support each other. The named ten parameters together with the angular coordinates ($\alpha$, $\delta$) and the proper motion ($\mu_\alpha$, $\mu_\delta$) of the system completely determine the dynamics of the system in the two-body approximation, including the masses of the two stars. In particular, the RVs together with the inclination angle $i$ give the absolute size $a$ of the orbit, its ratio to $\bar{a}$ is the distance $D$ of the system, and with Kepler's 3rd law we can also calculate the mass of the system.

Measurements of each observable are organized in separate input datasets, the zero points of which are determined in the {\sc Fotel} solution: for RVs the zero point is the $\gamma$-velocity, for astrometric separations it is the semimajor axis of the inner orbit, and for astrometric PAs it is the PA of the ascending node. Additional overall weights can be applied to balance the influence of each dataset on the final solution.  For more details about the main code, we refer to the {\sc Fotel} user guide \citep{2004PAICz..92....1H}. 

{\sc Fotel} can solve also the RV curves of a hierarchical triple system such as $\nu$~Gem. This means that the light-time effect and the motion of the center of mass of the inner binary due to the third component are taken into account. However, in the original version of {\sc Fotel}, the astrometry, i.e., the fit of angular distances and position angles, had been devised for the inner orbit only. We thus developed and carefully tested a new generalized version of {\sc Fotel} capable of simultaneously fitting the astrometry of both the inner (Ab relative to Aa) and outer orbit (B relative to Aa or to the center of light of Aa+Ab). In the case when the position of B is measured relative to Aa (as in our OLBI measurements), this version takes into account the reflex 'wobble' motion (hereafter referred to only as wobble) of the outer orbit due to the motion of Aa with respect to the centre of mass of the inner pair \citep[cf., e.g.,][]{2017ApJ...838...54T}.

The masses calculated separately for the inner and outer orbit should satisfy the condition $M_{\mathrm{A}}\equiv M_{\mathrm{Aa+Ab}}=M_{\mathrm{Aa}}+M_{\mathrm{Ab}}$. Due to observational errors this will, in practice, not be perfectly true if all parameters of the outer and inner orbit are solved for independently. To get a fully self-consistent solution, we therefore developed an alternative version of {\sc Fotel} in which the semi-amplitude $K_{\mathrm{Aa+Ab}}$ of the radial velocity of the center of mass of the inner binary is not a free parameter but is derived from the condition
\begin{equation}\label{cond}
     \frac{K_{\mathrm{B}}(K_{\mathrm{B}}+K_{\mathrm{A}})^2}{(K_{\mathrm{Aa}}+K_{\mathrm{Ab}})^3}=\frac{P\sin^3i'}{P'\sin^3i}\left(\frac{1-e^2}{1-e'^2}\right)^{3/2}\; ,
 \end{equation}
 \newline
which is derived from the equivalency of masses for the two orbits, and where $P$, $i$, $e$, and $P'$, $i'$, $e'$ are the period, inclination, and eccentricity of the inner and outer orbits, respectively. Sect.\,\ref{sec:triple_results} presents the results.

\section{Results}
\label{sec:analysis}

\begin{figure}[t]
   \centering
   \includegraphics[width=0.5\textwidth]{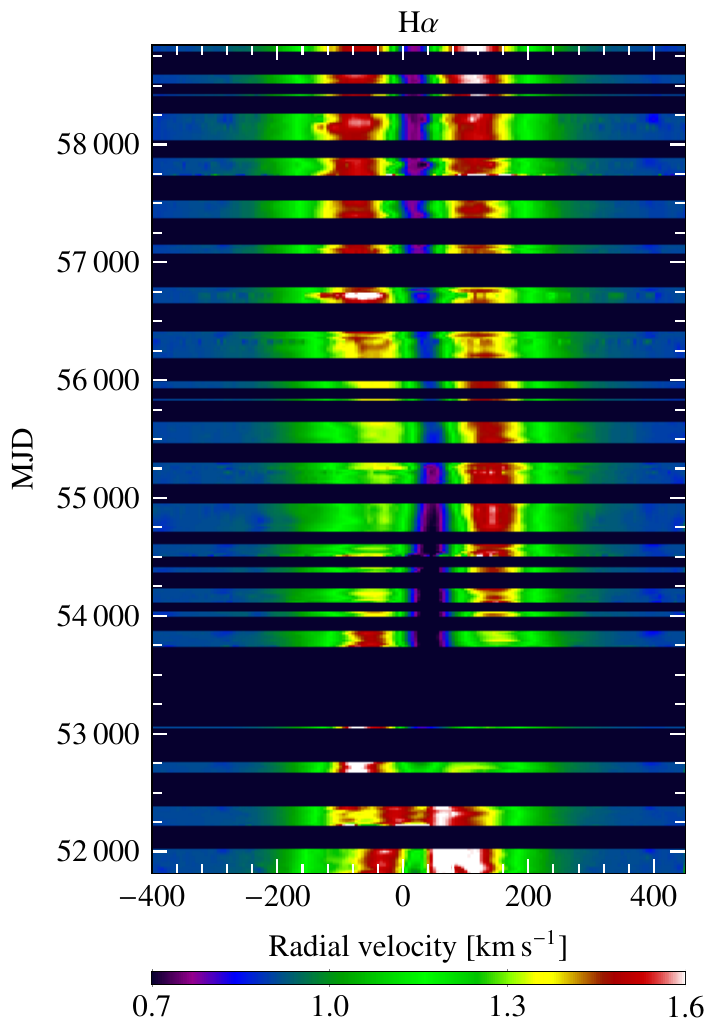}%
  
   \caption{\label{fig:Ha_linear} Dynamical spectrum of H$\alpha$ constructed from the complete set of observed spectra with the color scale representing the normalized flux.  The black horizontal stripes are due to gaps in the time coverage. The data happen to cover about one full outer orbital period of $\sim$7000\,d.  The central shell absorption traces the orbital motion but it is not present at all times.  Strong $V/R$ activity until MJD $\sim$57\,000 is also well visible.
   }
\end{figure}

\subsection{Spectral variability}
\label{sec:spec_variability}

For convenience, we recall here that $\nu$~Gem exhibits a composite spectrum of the three components in the system. The prominent photospheric spectrum moving with a $\sim53.7$\,day period belongs to one of the inner components with spectral type B6\,III \citep{2006A&A...459..137R}. The other prominent component of the composite spectrum is the Be star on the outer $\sim19$\,year-long orbit. The Be star spectrum includes emission lines, of which H$\alpha$ is the strongest. Because of the rotation of the disk, the H$\alpha$ profile has two peaks which have shown complex variability in the past decades, including strong $V/R$ variations and a triple-peaked profile (see Sect.~\ref{sec:literature} for more details). Simple visual inspection of our spectra does not reveal the second component of the inner system A.  

Prominent and sharp shell absorption features are regularly seen in high-resolution spectra in hydrogen Balmer lines and metal lines of \ion{Fe}{2}, \ion{Ca}{2}, and \ion{O}{1} but can be absent during some phases of the $V/R$ cycle.  They form in the circumstellar disk of the Be star.  The shell absorption can in both shape and position be subject to disk oscillations (manifested as $V/R$ variations) that could be intrinsic, caused by the gravitational influence of the inner pair or by an entirely putative, undetected closer companion of the Be star. Accordingly, some scatter in the RVs measured from shell absorption features with respect to the RV curve derived from the orbital solution is expected.  Nevertheless, with our full-cycle coverage, the orbital solution should still be valid (see Sect.~\ref{sec:triple_results}).

\begin{figure*}[t]
   \centering
   \includegraphics[width=\textwidth]{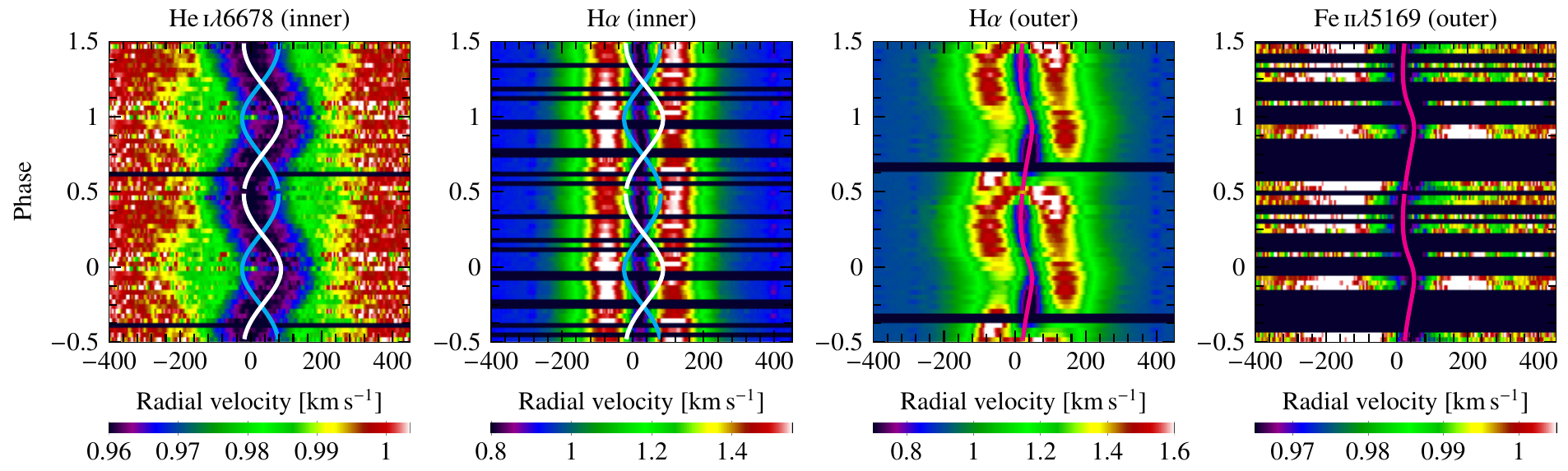}%
\caption{\label{fig:phased} Spectra of selected lines (as labeled) phased with inner (first and second panel from left) and outer (third and fourth panel) orbital periods, namely 53.772\,d and 19.159\,yr, respectively. The color scales (shown below each plot) represent the normalized flux. For the second panel, only profiles observed after MJD\,57000 were used at which time the effects of the $V/R$ density-wave pattern had mostly vanished (see Fig.\,\ref{fig:Ha_linear}). The deep absorption components in H$\alpha$ and \ion{Fe}{2} $\lambda$5169 are not photospheric but circumstellar shell lines of the Be star.  The combined solutions (RV+astrometry) from Table~\ref{tab3} (solution C4) are overplotted as continuous lines: in panels 1 and 2, the white line traces the motion of A2 (Aa in solution C4) and the cyan one that of A1 (Ab). 
In panels 3 and 4, the magenta line follows the motion of component B, and it can be seen that the emission lines from the disk of the Be star exhibit the same RV shifts.
}
\end{figure*}

The dynamical spectrum of H$\alpha$ constructed from all our combined-light spectroscopy is shown in Fig.~\ref{fig:Ha_linear}. From four H$\alpha$ profiles spanning $\sim$5 years and further data from the literature, \citet{1996A&AS..116..309H} estimate the cycle length of the $V/R$ variations to be about 5 years. However, their observation on Sept.~9, 1993 (MJD=49\,239) is very similar to that by \citet{2006A&A...459..137R} in the 2000/2001 season (MJD$\sim$52\,000), including an extended blue shoulder that is highly atypical for Be stars not showing the triple peak perturbation. 
The cycle length was, therefore, closer to $7.5$\,years, and was unrelated to the orbital periods before the variability faded in the following decade. The transition from $V/R<1$ to $V/R>1$ via a triple-peaked profile is seen at MJD $\sim$52\,300,
but after another 7.5-year cycle, at about MJD = 55\,000, the profile modulation has already weakened and then fully subsided and stabilized at $V=R$ by about MJD = 57\,000. Fig.\,\ref{fig:phased} (Panel 3) presents the same data phase-folded with the outer orbital period.

\begin{figure*}[t]
\gridline{\fig{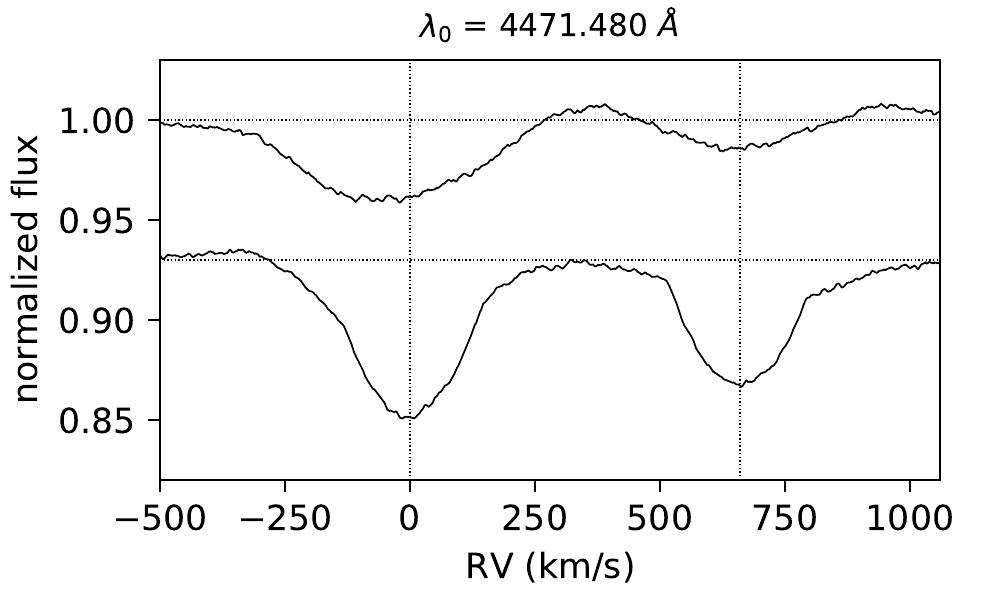}{0.5\textwidth}{}
          \fig{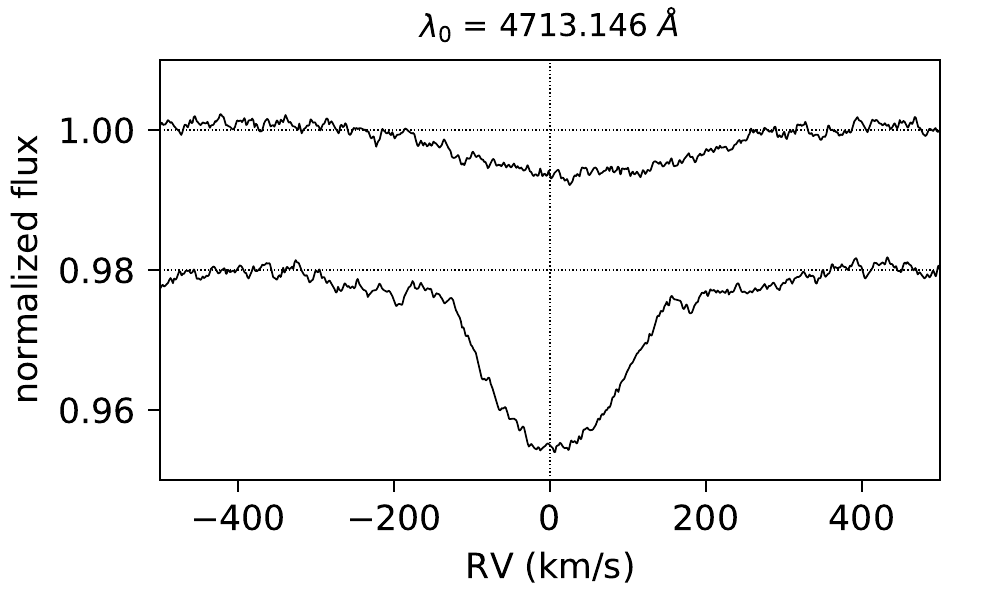}{0.5\textwidth}{}
         }
\gridline{
          \fig{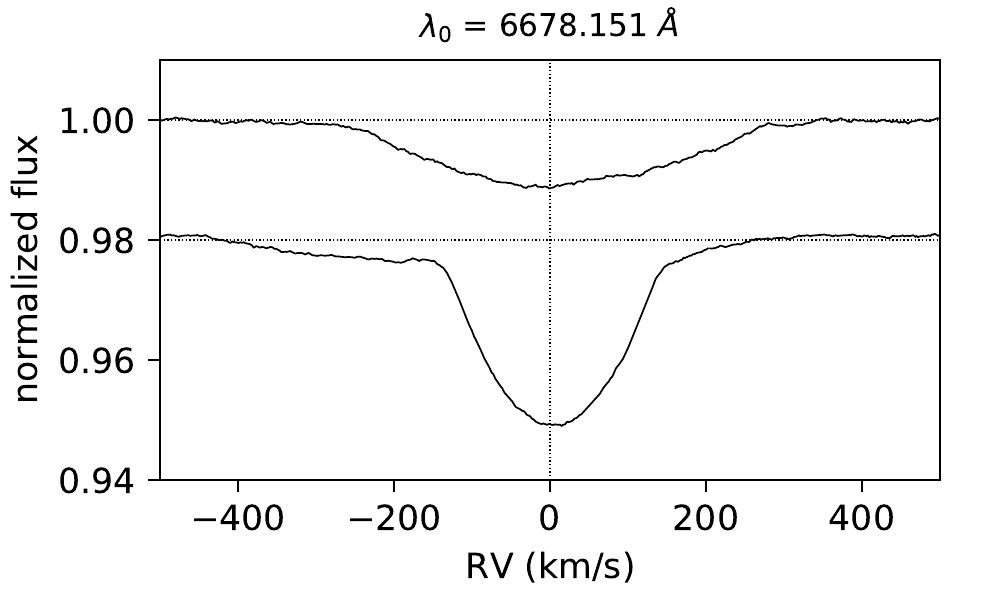}{0.5\textwidth}{}
          }
   \caption{\label{fig:disentangled_lines} Disentangled line profiles of \ion{He}{1}\,$\lambda4471$ and \ion{Mg}{2}\,$\lambda4481$ (upper left), \ion{He}{1}\,$\lambda4713$ (upper right) and \ion{He}{1}\,$\lambda6678$ (lower) belonging to the inner spectroscopic components A1 (broad-lined) and A2 (narrow-lined). The lines are normalized to the composite continuum, and those of  component A2 are vertically offset. The RVs have been corrected for the systemic velocity of $29.8$\,km/s (determined in Sect.~\ref{sec:triple_results}). As discussed in Sect.~\ref{sec:specmodel}, it is the spectroscopic component A2 that likely corresponds to the brighter astrometric component Aa.
   }
\end{figure*}

\begin{figure}[t]
   \centering
   \includegraphics[width=0.5\textwidth]{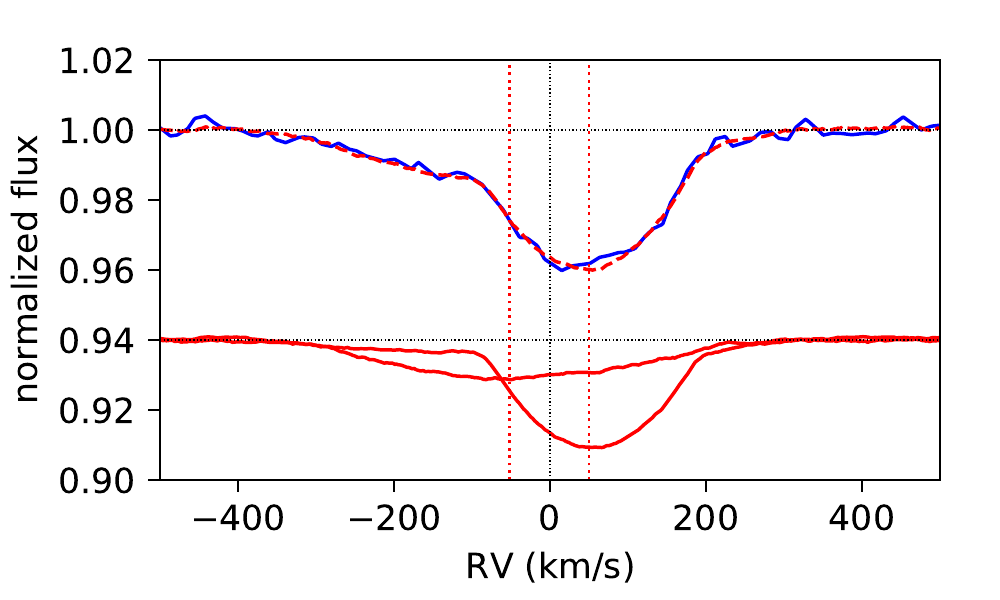}%
     \caption{\label{fig:korel_output} Example of an RV measurement performed by {\sc Korel} in the spectral region surrounding the \ion{He}{1} $\lambda 6678$ line. The line profile observed on MJD=53857.83 (blue) is overplotted with the best-fit combination of the disentangled profiles of the inner two components (dashed red). The individual disentangled and Doppler-shifted profiles (with very different rotational broadening) are plotted in solid red (shifted in flux by $-0.06$ continuum units). The dotted vertical red lines indicate the measured RVs of $-$51.7\,km~s$^{-1}$ and +50.0\,km~s$^{-1}$ for the broad-lined (A1) and narrow-lined (A2) inner component, respectively. The RVs have been corrected for the systemic velocity of $29.8$\,km~s$^{-1}$ (determined in Sect.~\ref{sec:triple_results}).
     }
\end{figure}

\subsection{Disentangled line profiles}
\label{sec:disentangling_results}

The {\sc Korel} code was used to disentangle the isolated line \ion{He}{1} $\lambda 6678$ using a total of 145 heterogeneous spectra from HEROS, ESPaDOns, and Ond\v{r}ejov coud\'{e} spectrographs (Table~\ref{tab:spectroscopy}). The procedure yielded disentangled line profiles of the inner components, their RVs, as well as spectroscopic orbital parameters (other lines were not used in this first step due to insufficient phase coverage of the outer orbit). For the disentangling procedure, each input spectrum was given a weight proportional to the square of the signal-to-noise ratio (SNR) of the input spectral region. The photospheric lines of component B (the Be star) remained undetected (see below), but the motion of the center of mass of the inner binary in the wide orbit was taken into account by including an invisible third component in the orbit model. 

The disentangled \ion{He}{1} $\lambda 6678$ line profiles of the inner components (normalized to the common continuum) are shown in the lower panel of Fig.~\ref{fig:disentangled_lines}, revealing that one component has much broader line profiles than the other. In order to have a unique notation for the inner spectroscopic components (as it is initially unclear which one is the brighter astrometric component Aa), in the following we will refer to the broad-lined component as A1 and the narrow-lined component as A2. The broadened line profile of component A1 indicates that it is a very fast rotator, which explains why the contribution from its broadened lines is not easily seen in combined-light spectra \citep[cf.][]{2006A&A...459..137R}. The $v\sin{i}$ derived from the disentangled profiles equals $260 \pm 20$\,\kms{} and $140 \pm 10$\,\kms\ for the broad-lined and narrow-lined components A1 and A2, respectively. The inner-component RVs measured by {\sc Korel} (see Fig.~\ref{fig:korel_output}) are used in the subsequent combined orbital analysis (Sect.~\ref{sec:triple_results}). The orbital parameters derived from the \ion{He}{1} $\lambda 6678$ line are given as spectroscopic solution S1 in Table~\ref{tab1}. The combined-light dynamical spectrum of the \ion{He}{1} $\lambda 6678$ line phased with the inner orbital period is presented in the left panel of Fig.~\ref{fig:phased}.

Additional line profiles were disentangled with the orbital parameters fixed according to solution S1 from the \ion{He}{1} $\lambda 6678$ line. The line profiles of \ion{He}{1}\,$\lambda4471$ and \ion{Mg}{2}\,$\lambda4481$ (disentangled as one spectral region), and \ion{He}{1} $\lambda 4713$ lines (all covered by 43 spectra) are shown in the upper two panels of Fig.~\ref{fig:disentangled_lines}. Using parameters from the combined orbital solutions (see next section) to disentangle the three spectral regions (Fig.~\ref{fig:disentangled_lines}) does not change the resulting profiles beyond very small details. 

In all aforementioned spectral regions, the disentangling did not produce photospheric line profiles from the Be star (component B). This is not surprising because the shell lines demonstrate that the Be star (an extremely rapid rotator) is viewed from a roughly equatorial perspective. Therefore, we used the shell lines originating in the Be star disk as substitute tracers of the Be star's RV curve. The shell lines are readily visible in the H$\alpha$ line, for which we have the same phase coverage as for the \ion{He}{1} $\lambda 6678$ line. However, the H$\alpha$ complex cannot be reliably disentangled due to the variability of the emission on aperiodic and non-orbital timescales. In order to constrain the orbital motion of component B for the full orbital solution (see the next subsection) we derived RVs in a classical way by fitting a single Gaussian to the H$\alpha$ shell absorption. The combined-light dynamical spectra of the H$\alpha$ and \ion{Fe}{2} $\lambda 5169$ lines (the latter is another line with a clear shell absorption) phased with the outer orbital period are shown in the right panels of Fig.~\ref{fig:phased}.

\begin{figure*}[t]
\gridline{\fig{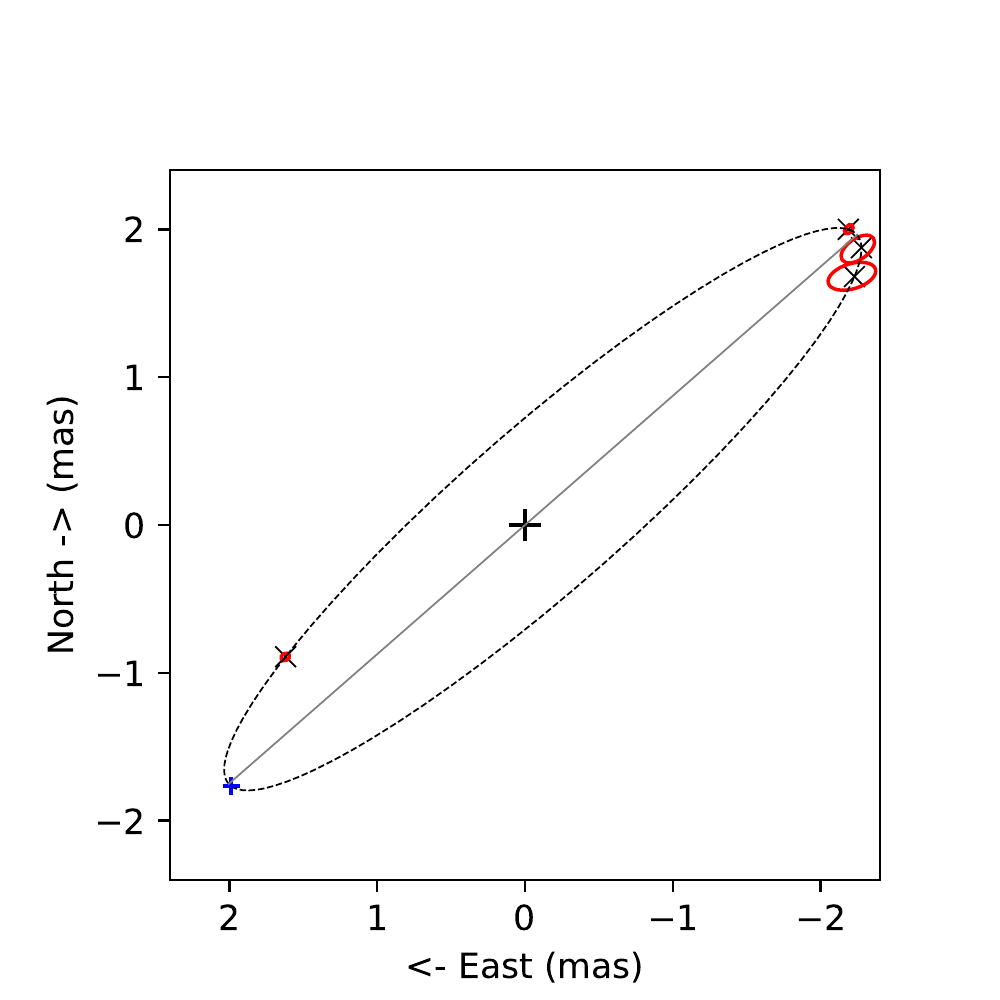}{0.45\textwidth}{}
          \fig{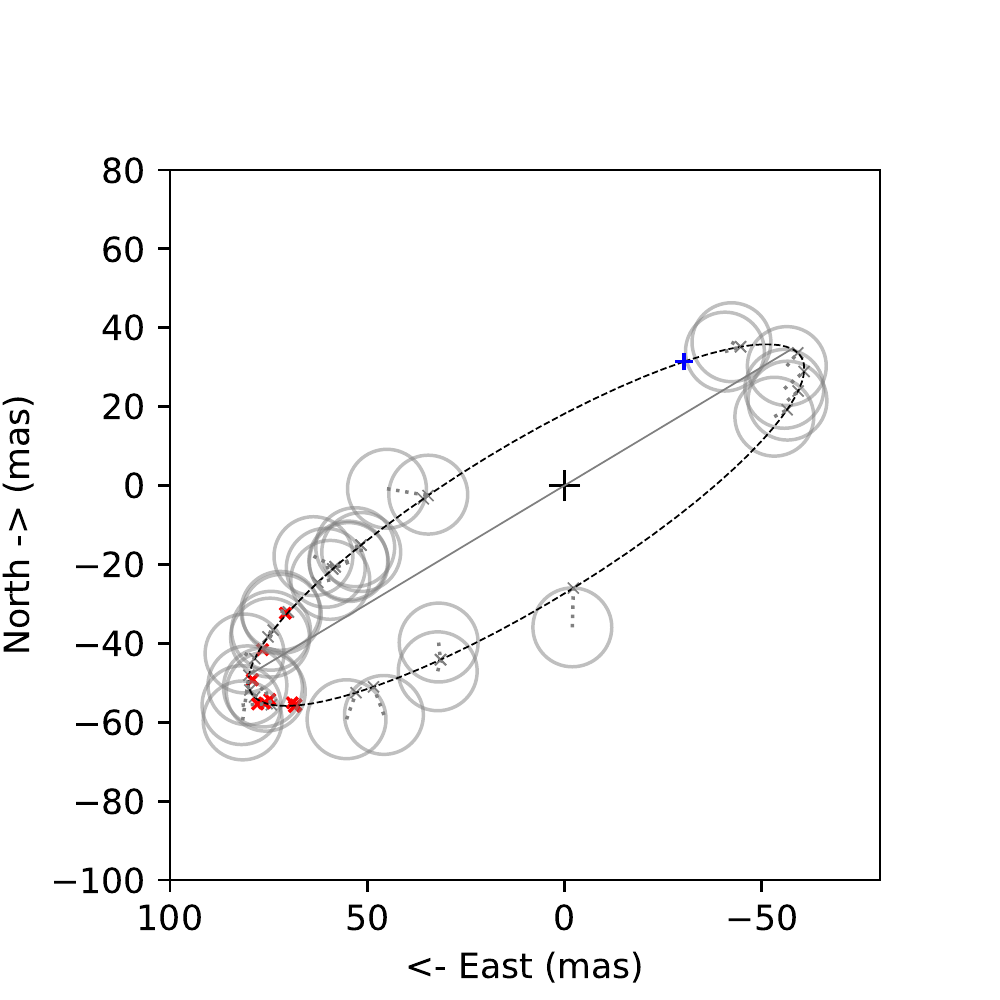}{0.45\textwidth}{}
         }
\caption{\label{fig:orbits} Astrometric orbits from solution C4 (Table\,\ref{tab3}). \textit{Left:} The Aa+Ab relative orbit (dashed line) with the brighter Aa component at the center. The MIRC/MIRC-X measurements are plotted as red 3-$\sigma$ error ellipses with the corresponding positions on the calculated orbit marked by black crosses. The node line is shown as a grey line and the periastron position appears as a blue plus sign. \textit{Right:} Same as left panel, but for the Aa+B relative orbit with the Aa component again at the center. The symbol sizes of the MIRC/MIRC-X measurements roughly correspond to the maximum wobble caused by the Aa motion around the inner center of mass. The WDS speckle measurements are shown as grey circles with a radius of 5\,mas, which is their typical error. The corresponding positions on the calculated orbit are marked by grey crosses and are connected to the measured positions (centers of the grey circles) by dotted grey lines.}
\end{figure*}

\begin{figure*}[t]
\gridline{\fig{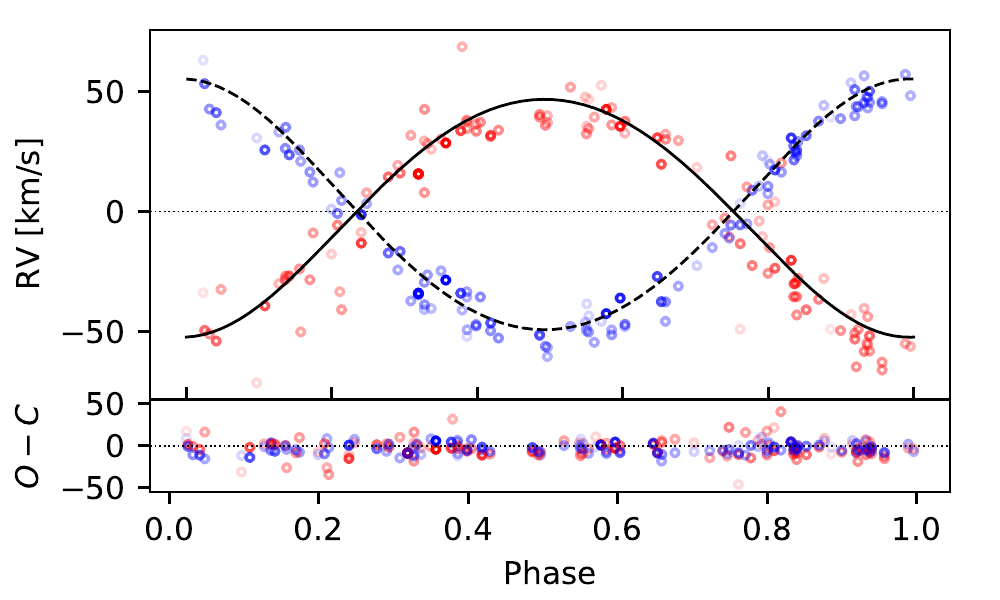}{0.5\textwidth}{}
          \fig{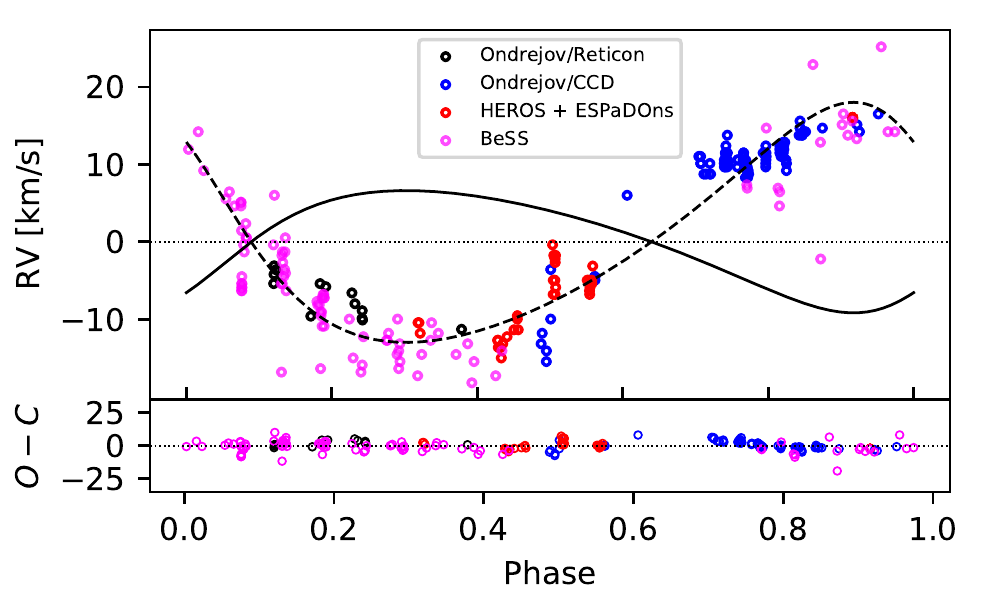}{0.5\textwidth}{}
         }
\caption{
\label{fig:RV_curves} RV curves from solution C4 (Table\,\ref{tab3}). \textit{Left:} RV curves of the narrow-lined component A2 (solid line) and the broad-lined component A1 (dashed line) in $\nu$~Gem with RVs measured by disentangling the \ion{He}{1}\,$\lambda6678$ (circles) line (see Sect.~\ref{sec:disentangling_results}). Measured RVs of the A1 component (red symbols) show a larger scatter, which is expected owing to its lines being much broader than those of the A2 component (blue symbols). The transparency of the symbols is scaled to represent the SNR of the input spectra (high transparency indicates low SNR). \textit{Right:} RV curve of the component B (dashed line) and the center of mass of the inner pair Aa+Ab (solid line). As a proxy for the RV of the Be star (component B), measurements of the H$\alpha$ shell absorption are used and plotted as circles, with the amateur spectra (BeSS) exhibiting a larger scatter compared to the other observations. The $O-C$ residuals of both curves are shown on a compressed scale in the lower plots. Both curves were corrected for the systemic velocity of $29.8$\,km/s (see Table~\ref{tab3}). 
}
\end{figure*}

\subsection{Orbital solutions of the triple system}
\label{sec:triple_results}

For the combined (spectroscopic + astrometric) orbital solution of $\nu$~Gem, we have seven observables available, namely one set of RVs for each of the three stars -- for the inner components A1 and A2 measured by {\sc Korel} and for component B measured by Gaussian fitting of the H$\alpha$ shell absorption -- and two sets of angular separations and position angles, respectively, for the inner (Aa/Ab) and outer (Aa/B) orbits. To account for the different precision of speckle and optical interferometry (the latter is three orders of magnitude more precise), we split the astrometry into separate datasets, making it a total of nine datasets used in the solution.  Despite the low accuracy of the speckle data, they are useful for constraining the outer orbit because of their good phase coverage. To make sure that each dataset has a similar influence on the final solution, we adjusted the overall weight of each one such that they contribute approximately the same amount to the sum of the residuals $\Sigma{(O-C)^2}$. As a final check of our combined solutions, we used the previously converged parameters as the starting values and ran {\sc Fotel} without including the archival speckle interferometry, which resulted in essentially identical solutions. 

As for the individual measurements within the observational datasets, each RV measurement by {\sc Korel} was given a weight proportional to the square of the SNR of the input spectral region, while each measurement of the H$\alpha$ shell absorption was weighted uniformly, except for the measurements using amateur spectra, which were given half the weight of the RVs derived from professional spectra. The astrometric positions measured by MIRC/MIRC-X were weighted in inverse proportion to the squares of the semimajor axes of the corresponding error ellipses (see Tables~\ref{tab:astro-mirc-inner} and \ref{tab:astro-mirc-outer}), while to each astrometric position taken from speckle measurements we assigned a weight corresponding to a 1-$\sigma$ error of 5\,mas because individual uncertainties are not in all cases available.

The application of a least-squares fit to observational data implicitly assumes that the free parameters to be fitted are overdetermined by the data. In our case, however, we have just four astrometric measurements of the inner orbit Aa/Ab, i.e., eight values from which seven parameters should be determined. This means that the data can be fitted relatively precisely so that small residuals would mimic an unrealistically high accuracy. We thus replaced each observed data point of the inner orbit astrometry by four 'ghost'-points at the vertices and co-vertices of their error ellipses to get a more realistic estimate of the errors of the data and the weight appropriate for each dataset.

We initialized the {\sc Fotel} calculations by using the RVs of the inner components (measured by {\sc Korel}, see Sect.~\ref{sec:disentangling_results}) to produce spectroscopic orbital solution S2 (Table~\ref{tab1}). This solution differs from S1 because the procedures used by {\sc Fotel} and {\sc Korel} are not equivalent (see Sect.~\ref{sec:methodology}); the difference is particularly significant for the outer orbit which is poorly constrained by the spectral lines of the inner pair (as we have not yet included RVs of component B). By including the shell-line RVs of component B in the orbital analysis, we arrived at spectroscopic solution S3 (Table~\ref{tab1}), which constrains the outer orbit much better and also yields the velocity semi-amplitude of component B $K_\mathrm{B}$, the mass ratio for the outer orbit $q'$, and the systemic velocity $\gamma$. For parameters that were not directly converged by {\sc Fotel} (such as for instance $K_\mathrm{A2}$ in solutions S2 and S3), the uncertainties were propagated while taking into account the correlation coefficients between the converged parameters (which are part of the {\sc Fotel} output). In Table~\ref{tab1}, the purely spectroscopic solutions S1--S3 are juxtaposed to the single-lined binary solution of \citet{2006A&A...459..137R}. 

For comparison with the purely astrometric orbital solution of \citet{2021AJ....161...40G}, we used only the astrometric datasets to derive solution A, which is compared to the \citet{2021AJ....161...40G} solution in Table~\ref{tab2}. The results show a good mutual agreement and similar residuals of the astrometric measurements: the median residuals for the Aa/Ab and Aa/B MIRC/MIRC-X astrometry for solution A are $6.1$\,$\mu$as and $97$\,$\mu$as, respectively, while for the \citet{2021AJ....161...40G} solution we get $10.2$\,$\mu$as and $106$\,$\mu$as, respectively.

For binaries, interferometric data suffer a fundamental 180\degree\ ambiguity in the longitude of the periastron, $\omega$.  In the case of the inner binary in $\nu$~Gem, this means that cross-identifying the brighter and the fainter components Aa and Ab distinguished by interferometry (Sect.\,\ref{sec:OLBI}) with the broad-lined and narrow-lined components A1 and A2 distinguished by spectral disentangling (Sect.\,\ref{sec:disentangling_results}) is initially arbitrary. For the combined (spectroscopy + astrometry) solutions C1-C4 presented below, we consider both possible options for the cross-identification of the spectroscopic and astrometric components, i.e., either Aa = A1 (the brighter star is the broad-lined one), or Aa = A2 (the brighter star has the narrower lines). As will be shown in Sect.~\ref{sec:specmodel}, the latter option is preferred based on the analysis of the disentangled line profiles.

The combined spectroscopy + astrometry solutions, which entail the complete description of the triple system including component masses and the distance, are listed in Table~\ref{tab3}. Solution C1 represents the complete solution with all orbital parameters converged, when assuming that Aa = A1.  While it provides a complete description of both orbits (Aa/Ab and Aa+Ab/B), the sum of $M_\mathrm{Aa}$ and $M_\mathrm{Ab}$ from the inner orbit solution does not exactly equal $M_\mathrm{A}$ from the outer orbit solution. To ensure the consistency of the masses, the condition of Eq.~\ref{cond} has to be enforced, so that $K_\mathrm{A}$ is not a free parameter. This results in the physically consistent solution C2, which is very similar to C1. The resulting orbits have a mutual inclination amounting to $i_\mathrm{r} = (152.9\pm0.6)${\degree} and are counter-rotating.

Solutions C3 and C4 are analogous to C1 and C2 but assume that Aa = A2. This was done by swapping the assignment of the inner-component RVs to obtain $M_\mathrm{Ab} > M_\mathrm{Aa}$, and therefore $q>1$. But as can be seen in Table~\ref{tab3}, in the {\sc Fotel} solution C4 (with enforced equality of masses), the inner mass ratio $q$ still converges to slightly below $1.0$. This is, however, not entirely unexpected, as the purely astrometric solution (A) results in $q<1.0$ ($M_\mathrm{Aa} > M_\mathrm{Ab}$), and swapping the inner RVs does not appear to be sufficient to reverse this.  On the other hand, the inner mass ratio is clearly close to unity, and when taking into account the associated errors, $q \simeq 1$ for both self-consistent solutions C2 and C4. This suggests that either scenario -- $q\gtrsim1$ or $q\lesssim1$ -- is acceptable in terms of agreement with the available orbital data.  In fact, since both C2 and C4 indicate that $q \simeq 1$, it follows that $M_\mathrm{Aa} \simeq M_\mathrm{Ab}$ to within the error margins. The mutual inclination of the two orbits resulting from solution C4 is $i_\mathrm{r} = (10.3\pm0.6)${\degree}, and the two orbits have the same direction. Solution C3, which would be analogous to solution C1 but with swapped inner components, results in inconsistent masses calculated from the inner and outer orbits, and the condition of Eq.~\ref{cond} in this case has to be enforced to obtain a physically meaningful solution.  For this reason, C3 is not included in Table~\ref{tab3}.

\begin{figure}[t]
   \centering
   \includegraphics[width=0.5\textwidth]{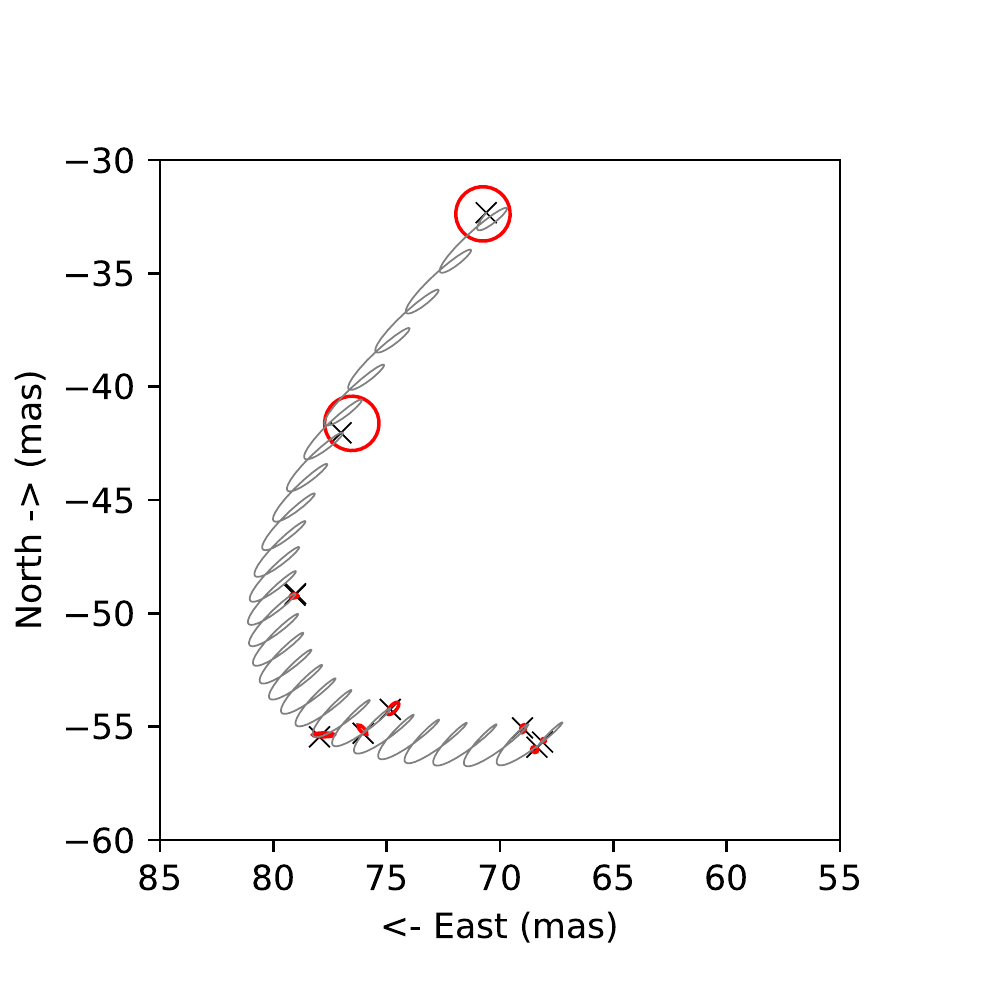}%
     \caption{\label{fig:outer_orbit_detail} Detail of the outer orbit from solution C4 (Table\,\ref{tab3}) including the wobble exerted by the inner binary. The interferometric positions are plotted as 3-$\sigma$ error ellipses (red) and the corresponding positions on the orbit are marked by black crosses. The median residual of these ten points is $\sim130$\,$\mu$as.
     }
\end{figure}

\begin{figure*}[t]
\gridline{\fig{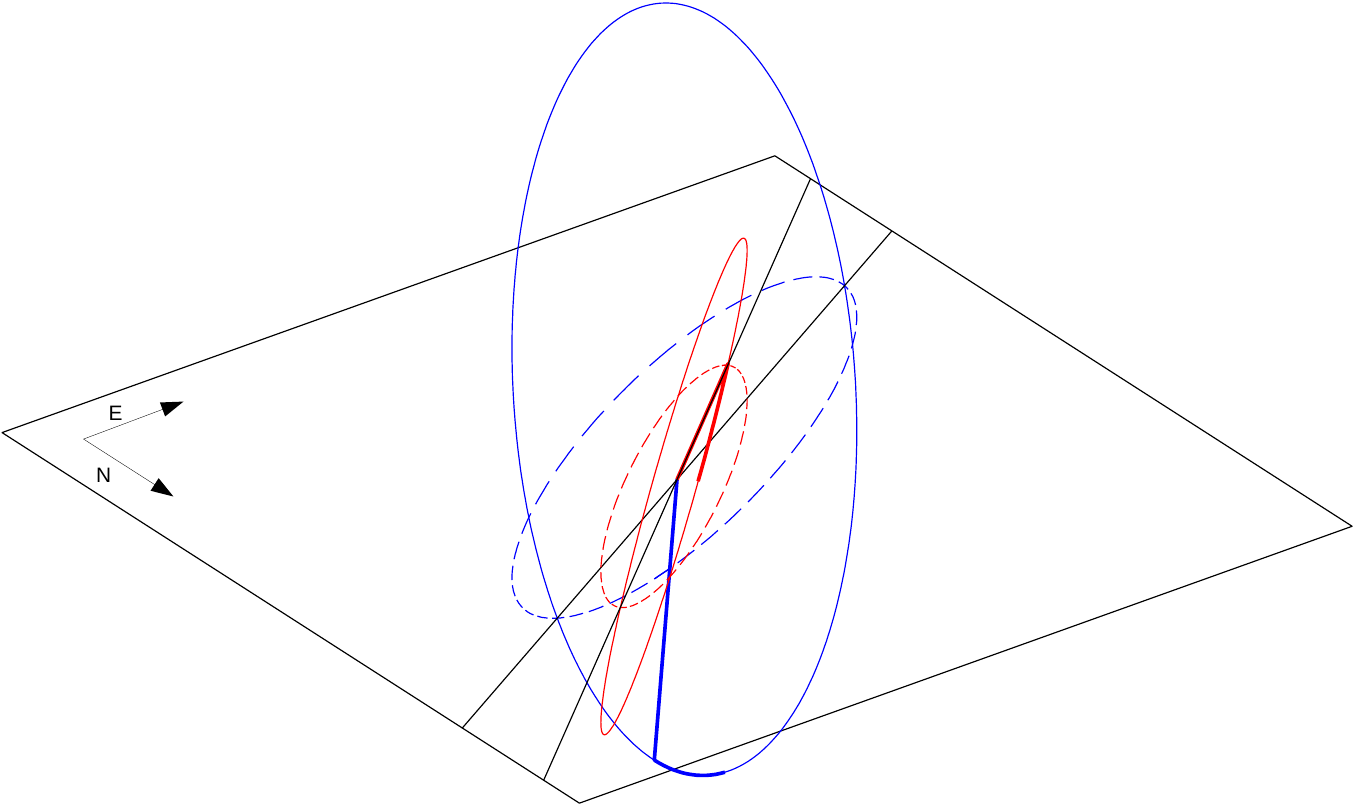}{0.5\textwidth}{}
          \fig{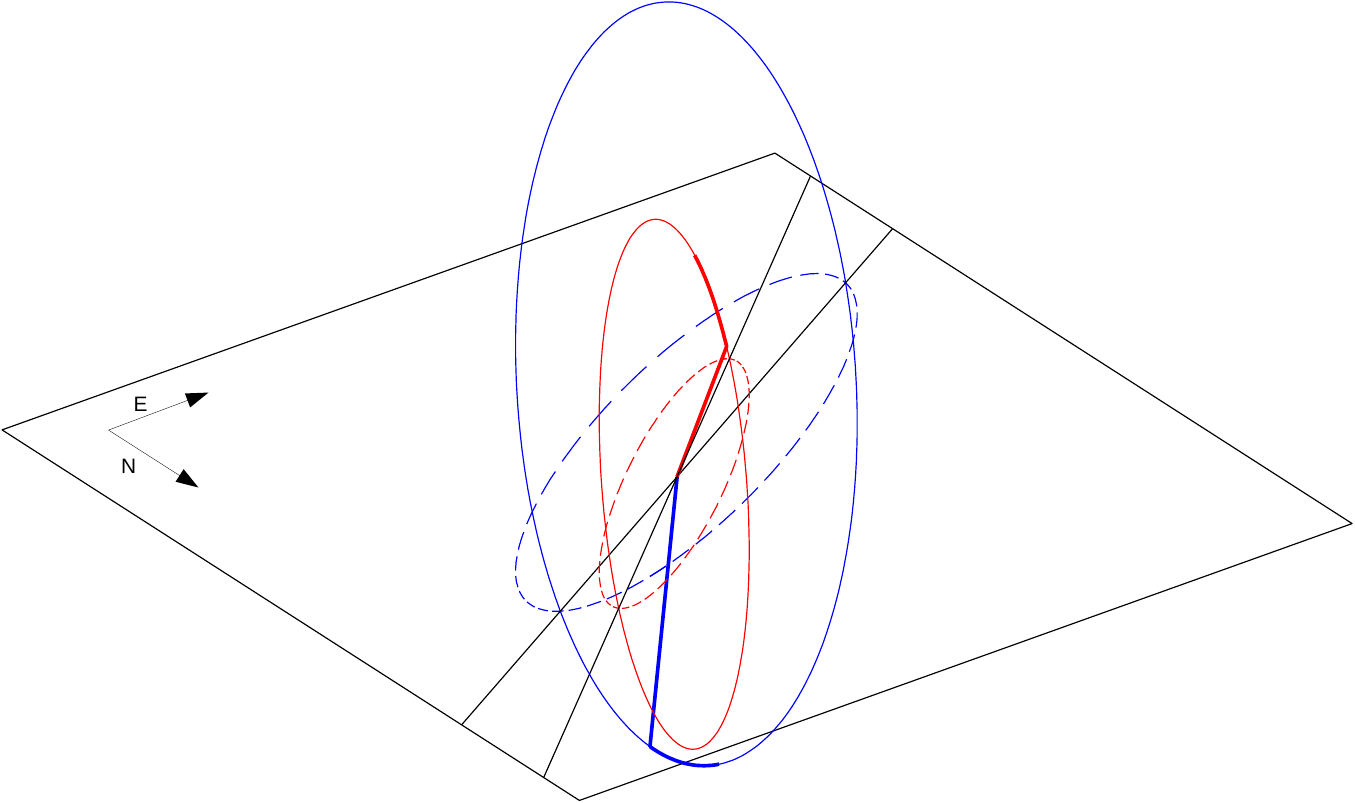}{0.5\textwidth}{}
         }
\caption{\label{fig:orbit_orientation} 3-D configuration of $\nu$~Gem in space for solutions C2 (\textit{left}, the counter-rotating case in which the brighter Aa is the broad-lined A1) and C4 (\textit{right}, the co-directional case in which Aa is the narrow-lined A2). The black square defines the plane of the sky that is viewed from the top; the cardinal directions are labeled. The two black lines mark the lines of nodes of the two orbits. Solid ellipses delineate the outer (blue) and the inner (red; 20$\times$ enlarged) orbit, respectively.  Thick colored lines connect the origin with the periastra.  The thick segments in either orbit mark the first one-twelfth of the respective orbit after periastron passage and so indicate the direction of the orbital motion. The projections of the orbits onto the plane of the sky are dashed with a period of 10\,{\degree} in mean anomaly. 
}
\end{figure*}

We note that both self-consistent solutions C2 and C4 fit the orbital data equally well. However, as will be shown from the analysis of the disentangled line profiles in Sect.~\ref{sec:specmodel}, C4 is considered as the preferred solution.  We recall that, in solution C4, Aa is the narrow-lined spectroscopic component A2, and Ab is the broad-lined component A1, while their masses are not significantly different. The astrometric orbits corresponding to solution C4 are plotted in Fig.~\ref{fig:orbits} while the RV curves are presented in Fig.~\ref{fig:RV_curves}.  A zoom-in into part of the outer orbit and its agreement with the interferometric data including the wobble exerted by the inner binary is shown in Fig.~\ref{fig:outer_orbit_detail}. The orientation of the $\nu$~Gem orbit in space (for both C2 and C4) is depicted in Fig.~\ref{fig:orbit_orientation}. 

The distances derived from our orbital solutions -- $181.1\pm10.8$\,pc and $185.5\pm4.3$ from solutions C2 and C4, respectively -- are somewhat higher than the revised Hipparcos distance of $167\pm8$\,pc \citep{2007A&A...474..653V}, while Gaia distances are unreliable for objects as bright as $\nu$~Gem. The discrepancy is, however, not unexpected, as the published parallaxes do not take into account the orbital motions due to multiplicity.  Accordingly, the distance derived here from the full orbital solution should be more reliable.

\begin{table}
\centering
\caption{Purely spectroscopic solutions S1-S3 of the orbital parameters of $\nu$~Gem (cf. Sects. \ref{sec:disentangling_results} and \ref{sec:triple_results}) compared to the previous spectroscopic solution of \citet{2006A&A...459..137R}.}\label{tab1}
{\small
\begin{tabular}{lcccc}
                       &S1              &S2             &S3    & R2006$^\mathrm{a}$          \\ \hline
\multicolumn{5}{c}{Inner orbit}\\ \hline                                                                                                                      
$P$ [d]	               &$53.761\pm.003$	&$53.763\pm.005$&$53.762\pm.004$ & $53.731\pm0.017$ \\
$T$ [MJD]	       &$51006.1\pm0.7$	&$51006.5\pm0.9$&$51006.6\pm0.9$ & $51004.7\pm4.2$\\
$e$		       &$0.079\pm.007$ 	&$0.077\pm.008$	&$0.075\pm.007$	& $0.11\pm 0.5$ \\
$\omega$ [\degree]     &$146.2\pm4.5$	&$148.7\pm6.0$	&$149.2\pm6.1$ & $315\pm 29$	 \\
$K_\mathrm{A1}$ [km/s] &$48.5\pm1.6$    &$48.4\pm1.4$   &$48.3\pm1.5$ & 	 \\
$K_\mathrm{A2}$ [km/s] &$51.7\pm0.5$    &$52.0\pm0.6$   &$52.0\pm0.8$ & $38\pm 8$	 \\
$q$		       &$0.94\pm0.03$	&$0.93\pm0.03$  &$0.93\pm0.03$ & 	 \\
\hline                                                                                                                                                        
\multicolumn{4}{c}{Outer orbit}\\ \hline                                                                                                                      
$P'$ [d]	       &$6402\pm719$	&$6375\pm16000$ &$7160.7\pm10.5$  & \\
$T'$ [MJD]	       &$49224\pm517$	&$49349\pm1170$ &$49081.3\pm22.9$ & \\
$e'$		       &$0.381\pm0.045$ &$0.305\pm0.466$&$0.100\pm0.001$ & \\
$\omega'$ [\degree]&$252.0\pm8.9$	&$253.0\pm84.5$ &$253.1\pm0.8$ &   \\
$\gamma$ [km/s]    &            	&               &$29.4\pm0.1$ &   \\
$K_\mathrm{A}$ [km/s]  &$8.1\pm0.5$	&$8.0\pm28.5$   &$7.1\pm0.5$ &     \\
$K_\mathrm{B}$ [km/s]  &            &               &$14.1\pm0.1$ &     \\
$q'$		       &               	&               &$0.50\pm0.04$ &  \\
\end{tabular}
 } 
 \tablecomments{a - \citet{2006A&A...459..137R} only detected lines of the sharp-lined A2 component (which they assumed to be the primary component), which is why their $\omega$ differs from our solutions by about 180\degree.}
\end{table}   

 \begin{table}
\centering
\caption{Purely astrometric solution A of the orbital parameters of $\nu$~Gem (cf. Sects. \ref{sec:disentangling_results} and \ref{sec:triple_results}) compared to the also purely astrometric solution of \citet[][wobble + visual fit for the inner orbit]{2021AJ....161...40G}.}\label{tab2}
{\small
\begin{tabular}{lcc}
                       &A              & G2021          \\ \hline
\multicolumn{3}{c}{Inner orbit}\\ \hline
$P$ [d]	              &$53.743\pm.007$  &$53.7276\pm0.0066$\\
$T$ [MJD]	        &$58487.56\pm0.02$  &$58488.6\pm2.7$ \\
$e$		               &$0.041\pm.004$  &$0.0303\pm0.004$ \\
$\omega$ [\degree] &$18.4\pm0.1^\mathrm{a}$ &$26\pm18^\mathrm{a}$ \\
$q$		       	        &$0.983\pm0.026$&$0.895\pm0.028$\\
$i$ [\degree]	         &$79.5\pm0.5$  &$79.76\pm0.33$ \\
$\Omega$ [\degree]      &$131.1\pm1.1^\mathrm{a}$  &$131.17\pm0.16^\mathrm{a}$ \\
$\bar{a}$ [mas]	  	  &$2.863\pm0.013$  &$2.895\pm0.019$  \\
\hline                                                      
\multicolumn{3}{c}{Outer orbit}\\ \hline                     
$P'$ [d]	       &$6978.2\pm6.3$  &$6985\pm18$  \\
$T'$ [MJD]	       &$55847.6\pm32.7$&$55939\pm74$ \\
$e'$		          &$0.25\pm0.01$&$0.28\pm0.01$\\
$\omega'$ [\degree]  &$229.4\pm1.4$ &$233\pm3$\\
$i'$ [\degree]	     &$76.00\pm0.15$&$75.92\pm0.15$\\
$\Omega'$ [\degree]  &$120.8\pm1.2$ &$120.19\pm0.28$\\
$\bar{a}'$ [mas]      &$83.1\pm1.2$ &$83.12\pm0.59$ \\
\end{tabular}
 } 
\tablecomments{a -There is a 180{\degree} ambiguity in the purely astrometric solution. Values with $\Omega<180$\degree are reported.}
\end{table}

\begin{table}
\centering
\caption{Combined spectroscopic and astrometric solutions C1, C2, and C4 of the  orbital parameters of $\nu$~Gem (cf.\ Sects. \ref{sec:disentangling_results} and \ref{sec:triple_results}).
}\label{tab3}
{\small
\begin{tabular}{lccc}
                          &C1              &C2      &C4 \\ \hline
\multicolumn{4}{c}{Inner orbit}\\ \hline    
$P$ [d]	        &$53.7713\pm0.0024$&$53.7713\pm0.0024$&$53.7722\pm0.0008$\\
$T$ [MJD]	          &$51011.2\pm0.7$ &$51011.2\pm0.6$&$51011.8\pm0.1$ \\
$e$		              &$0.058\pm0.007$ &$0.057\pm0.005$&$0.056\pm.003$  \\
$\omega$ [\degree]      &$182.7\pm4.6$ &$182.5\pm3.4$  &$6.7\pm2.0$     \\
$K_\mathrm{Aa}$ [km/s]  &$49.1\pm3.8$  &$49.4\pm3.3$   &$51.6\pm0.6$    \\
$K_\mathrm{Ab}$ [km/s]  &$52.2\pm1.4$  &$52.1\pm0.6$   &$52.5\pm1.1$    \\
$q$		       	        &$0.94\pm0.09$ &$0.95\pm0.06$  &$0.98\pm0.03$   \\
$i$ [\degree]	        &$78.8\pm0.5$  &$78.8\pm0.3$   &$78.9\pm0.2$    \\
$\Omega$ [\degree]      &$311.1\pm0.1$ &$311.1\pm0.1$  &$131.1\pm0.1$   \\
$\bar{a}$ [mas]	  	    &$2.82\pm0.07$ &$2.82\pm0.05$  &$2.82\pm0.02$   \\
$M_\mathrm{Aa}$ [\Msun] &$3.14\pm0.26$ &$3.16\pm0.27$  &$3.34\pm0.15$   \\
$M_\mathrm{Ab}$ [\Msun]	&$2.95\pm0.44$ &$3.00\pm0.43$  &$3.28\pm0.10$   \\
$D$ [pc]	            &$180.6\pm10.0$ &$181.1\pm10.8$  &$185.5\pm4.3$   \\
\hline                                                                                                                                                        
\multicolumn{4}{c}{Outer orbit}\\ \hline                                                                                                                      
$P'$ [d]	          &$6990.4\pm6.3$  &$6990.9\pm6.2$  &$6977.3\pm6.1$  \\
$T'$ [MJD]	          &$48819.1\pm14.0$&$48818.8\pm14.0$&$48810.3\pm13.0$\\
$e'$		          &$0.242\pm0.002$ &$0.242\pm0.002$ &$0.241\pm0.002$ \\
$\omega'$ [\degree]   &$228.0\pm0.5$   &$228.0\pm0.5$   &$226.9\pm0.4$   \\
$\gamma_3$ [km/s]     &$29.9\pm0.1$    &$29.9\pm0.1$    &$29.8\pm0.1$    \\
$K_\mathrm{A}$ [km/s] &$7.9\pm0.6$     &$7.9\pm0.5$     &$8.0\pm0.1$     \\
$K_\mathrm{B}$ [km/s] &$15.5\pm0.1$    &$15.5\pm0.1$    &$15.9\pm0.1$    \\
$q'$		          &$0.51\pm0.04$   &$0.51\pm0.04$   &$0.50\pm0.01$   \\
$i'$ [\degree]	      &$76.0\pm0.1$    &$76.0\pm0.1$    &$75.9\pm0.2$    \\
$\Omega'$ [\degree]   &$121.0\pm0.1$   &$121.0\pm0.1$   &$121.0\pm0.1$   \\
$\bar{a}'$ [mas]      &$83.4\pm8.2$    &$83.0\pm6.4$    &$82.8\pm1.3$    \\
$M_\mathrm{A}$ [\Msun]&$6.17\pm0.35$   &$6.16\pm0.32$   &$6.62\pm0.16$   \\
$M_\mathrm{B}$ [\Msun]&$3.15\pm0.42$   &$3.14\pm0.36$   &$3.33\pm0.10$   \\
\end{tabular}
 } 
\end{table}      

\subsection{Basic parameters of the components derived from spectral modeling}
\label{sec:specmodel}

\begin{figure}[t]
   \centering
   \includegraphics[width=0.5\textwidth]{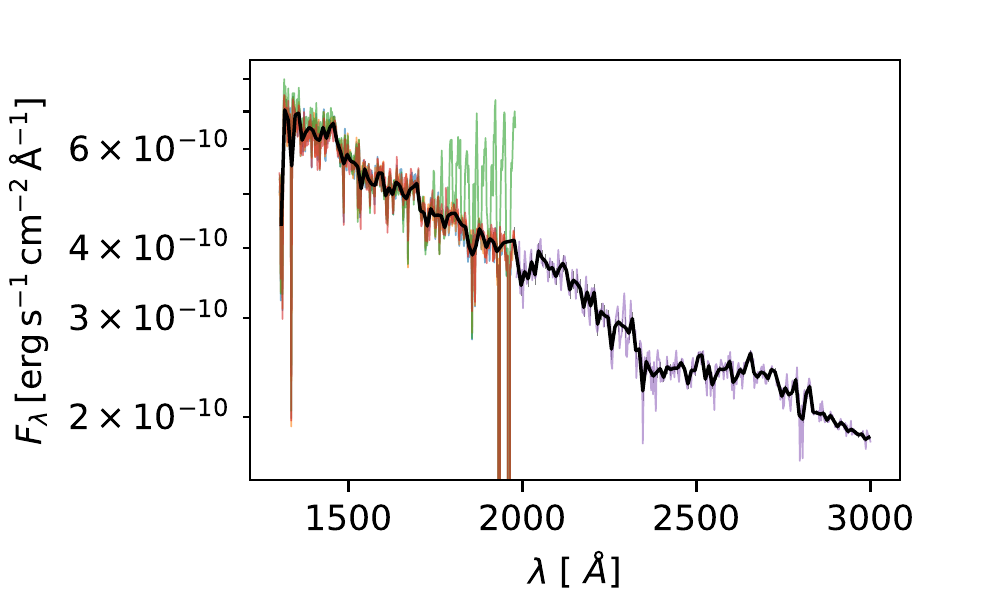}%
     \caption{\label{fig:iue_spec} Individual IUE spectra (colored) overplotted with their rebinned average (black). No interstellar absorption/redenning bump is evident.}
\end{figure}

A cursory inspection of the observed spectra, the UV spectral energy distribution (SED; Fig.~\ref{fig:iue_spec}), and the interferometric $H$-band flux ratios (Table~\ref{tab:rel_flux}) suggests that all three components of $\nu$~Gem are mid- to late-type B stars, which is in agreement with the masses derived from the combined orbital solutions (Table~\ref{tab3}).  Mid- to late-type B stars have very small color terms in the luminosity calibration, with even smaller variations, and Be stars in that range do not typically have a strong near-infrared flux excess.  Therefore, compared to all other uncertainties, a direct transfer of the $H$-band flux ratios to the $V$- and $R$-bands should not introduce any dominating systematic error. 

For a total $V$-band magnitude of $m_{V}=4.14$\,mag \citep{2002yCat.2237....0D}, the component magnitudes become $m_{V, \mathrm{Aa}} = 5.04$\,mag, $m_{V, \mathrm{Ab}} = 5.76$\,mag, and $m_{V, \mathrm{B}} = 5.31$\,mag, respectively. The interstellar reddening to $\nu$\,Gem is close to zero; \citet{1990MNRAS.247..407F}, for instance, determined $E(B-V)=1.35\times E(b-y) = -0.01\pm0.04$. This is in agreement with the complete absence of the $2200$\,\AA\ interstellar absorption/reddening bump \citep{2013AN....334.1107Z} in the IUE spectra (Fig.~\ref{fig:iue_spec}). Accordingly, the apparent magnitudes $m$ can be converted to absolute magnitudes $M$ using only the distance. With the distance of $185.5$\,pc from solution C4 (corresponding to a distance modulus of $\mu =5\log(185.5)-5=6.30$\,mag), this results in $M_{V,\rm Aa} = -1.30$\,mag, $M_{V, \mathrm{Ab}} = -0.58$\,mag, and $M_{V, \mathrm{B}} = -1.03$\,mag which are not extraordinary values for mid- to late-type B stars \citep{2006MNRAS.371..185W}. 

The disentangled line profiles of the spectroscopic inner components A1 and A2 (Fig.~\ref{fig:disentangled_lines}) are normalized to the common continuum and for spectral modeling they need to be scaled to the continuum of the individual components.  For this, we again assume that the flux ratios are the same as in the $H$-band (Table~\ref{tab:rel_flux}), while taking into account that either A1 or A2 can correspond to the brighter inner component Aa. It is important to note that while the shape of the disentangled spectral lines is rather robust, their actual flux scaling is less certain, as even with the flux ratios known from interferometry, the disentangled spectra suffer from inhomogeneous input spectra and the resulting uncertainties of the continuum normalization. Therefore, line parameters such as absolute equivalent widths (EWs) that are easily measured in single-component spectra can only be determined with large uncertainties in our case. 

Measuring the disentangled line EWs when assuming that Aa = A1 (corresponding to the combined solutions C1-C2) and comparing them with the calculations of \citet{1998A&A...330..306L} for main sequence B-type stars (luminosity classes V, IV, and III) reveals a discrepancy in the resulting effective temperatures of the inner components: while the spectra as well as the derived masses are consistent with mid- to late-type B stars for both inner components, the EWs point towards the Ab component being an early-type B star, or even a late-type O star.

The alternative option when Aa = A2 and Ab = A1 (corresponding to combined solutions C3-C4) shows a much better agreement with the derived temperatures and spectral types. Therefore, we use Aa = A2 and Ab = A1 as the preferred cross-identification of the spectroscopic and astrometric components in what follows. The measured EWs of the lines of the inner components for this option are listed in Table~\ref{tab:EWs}. The values are in this case consistent with effective temperatures in the range of 14--18\,kK, corresponding to mid to late B spectral type for both components. The rather wide spread of temperatures determined in this way probably reflects the uncertainty of the relative strengths of disentangled profiles when the absorption is at a level of only a few percent of the continuum (see Fig.~\ref{fig:disentangled_lines}).

\begin{deluxetable*}{cCCC}
\tablecaption{Equivalent widths of the disentangled \ion{He}{1} and continuum-scaled (according to Table~\ref{tab:rel_flux}) line profiles. \label{tab:EWs}}
\tablewidth{0pt}
\tablehead{
\colhead{component} & \colhead{EW \ion{He}{1}\,$\lambda4471$ [\AA]} & \colhead{EW \ion{He}{1}\,$\lambda4713$ [\AA]} & \colhead{EW \ion{He}{1}\,$\lambda6678$ [\AA]} }
\startdata
A1=Ab  & 1.06 & 0.14 & 0.40\\
A2=Aa & 0.68 & 0.21 & 0.35 \\
\enddata
\end{deluxetable*}

The broad-lined inner component A1 (Ab) is a very rapid rotator ($v \sin{i} = 260 \pm 20$\,\kms), while for component A2 (Aa) the rotation rate is also non-negligible ($v \sin{i} = 140 \pm 10$\,\kms). Component B, too, is a fast rotator, as classical Be stars such as component B generally rotate close to the critical value \citep{2013A&ARv..21...69R}.  The expected shallow spectral lines are in agreement with the nondetection of the Be star's photospheric spectrum by the disentangling procedure. In contrast to slowly and moderately rotating stars, the spectral features and the SED of rapid rotators depend on viewing angle and cannot be translated straightforwardly to evolutionary parameters \citep[see, e.g.,][for a discussion]{2004MNRAS.350..189T}. Most importantly, gravity darkening leads to a stellar latitude-dependent $T_\mathrm{eff}$ and $\log{g}$ \citep[Sect.~2.3.1 of][]{2013A&ARv..21...69R, 1924MNRAS..84..665V, 1963ApJ...138.1134C}.  Therefore, the usual calibrations in, e.g., $T_\mathrm{eff}$ of the observed spectral and SED features become more complex.

In an attempt to obtain more realistic component parameters that take into account the rapid rotation and gravity darkening, we employed the spectral synthesis code {\sc B4} \citep{2013MNRAS.429..177R, 2018A&A...609A.108S}, using synthetic spectra from the ATLAS9 LTE grid of model atmospheres \citep{1992IAUS..149..225K}. {\sc B4} produces the line profiles as well as the continuum, resulting in a perfectly normalized spectrum. The effective temperature derived by {\sc B4} represents the uniform temperature in the sense that it is the temperature at which a black body of the same surface area would have the same luminosity as the gravity-darkened star. For the gravity-darkening exponent $\beta$ \citep{1963ApJ...138.1134C} we adopt an intermediate value of 0.22 \citep[based on results from optical interferometry reviewed by][]{2012A&ARv..20...51V}.

The apparent spectrum of a rotating star is fully determined by five parameters, namely the mass, (polar) radius, luminosity (i.e., total energy output), rotation rate, and inclination of the object. For the $\nu$\,Gem system, the masses and the inclinations are available from the orbital solution and can, therefore, be fixed for the modeling.  We can furthermore assume that the rotational axes are aligned with the pertinent orbital axis so that the rotation rates can be derived from the measured $v\sin{i}$ values of the Aa and Ab components and from the modeled orbital inclination, while for the Be star near-critical rotation can be adopted. This leaves only the luminosity and the (polar) radius of each star as free parameters to be determined from the disentangled line profiles and interferometric flux ratios.

We used {\sc B4} to compute model line profiles of \ion{He}{1}\,$\lambda4471$,  \ion{Mg}{2}\,$\lambda4481$, \ion{He}{1}\,$\lambda4713$, and  \ion{He}{1}\,$\lambda6678$ for comparison with the disentangled lines. Reproducing the line-depth ratio of the \ion{He}{1}\,$\lambda4471$ and \ion{Mg}{2}\,$\lambda4481$ lines, which is used for the spectral classification of mid- to late-type B stars \citep[e.g.,][]{1992oasp.book.....G}, we arrive at the uniform $T_\mathrm{eff}$ of $\sim13$\,kK and $\sim16$\,kK for the narrow-lined (Aa) and broad-lined (Ab) component, respectively. However, as can be seen in Fig.~\ref{fig:disentangled_lines}, the disentangled profile of the broad-lined component has a continuum level slightly above unity around the \ion{Mg}{2} line, thus enhancing the line depth ratio for this component, while reducing it for the narrow-lined component. The temperatures of the inner components are therefore likely similar and after correcting for the uneven continuum level we arrive at effective temperatures of $\sim 14$\,kK and $\sim 15$\,kK for the Aa and Ab components, respectively.

Fitting the modeled line profiles of \ion{He}{1}\,$\lambda4713$ and \ion{He}{1}\,$\lambda6678$ directly to the disentangled and scaled profiles results in rather discrepant values of $T_\mathrm{eff}$ in the range of 15--17\,kK for component Aa and 15--18\,kK for component Ab. The \ion{He}{1}\,$\lambda6678$ line  gives a $T_\mathrm{eff}$ that is systematically higher by about 2\,kK.  Fitting the \ion{He}{1}\,$\lambda4471$ and \ion{Mg}{2}\,$\lambda4481$ lines, we find that the model profiles do not reach the depth of the disentangled profiles for any reasonable parameter combination. This failure could be explained by the fact that the outer Be star makes a non-negligible contribution to the observed absorption profiles (especially for \ion{He}{1}\,$\lambda4471$ / \ion{Mg}{2}\,$\lambda4481$ and \ion{He}{1}\,$\lambda6678$) that the disentangling apportions to the profiles of the inner components so that the lines appear deeper than they actually are. We note that this discrepancy gets much worse when assuming that Aa = A1, confirming the result from line EWs that the Aa = A2 option presents a better agreement with the disentangled lines. 

Given the spread of the temperature determinations, we conservatively adopt $T_\mathrm{eff} = 15 \pm 3$\, kK for both inner components Aa and Ab. Not having detected any photospheric lines from component B, we adopt $T_\mathrm{eff} = 12 \pm 3$\, kK according to the reported spectral type of B8III \citep{1997AJ....114..808M}. For more precise estimates, it is paramount to obtain new, high signal-to-noise spectra with good phase coverage to hopefully be able to disentangle also the photospheric lines of the Be star.

The line profiles described above are rather insensitive to different combinations of luminosities $L$ and polar radii $R_\mathrm{P}$ that result in the correct rotation rate (as determined by the measured $v\sin{i}$, mass, and inclination). We thus used {\sc B4} to also compute low-resolution $H$-band spectra and compared the resulting absolute flux levels with the measured interferometric flux ratio in order to resolve this degeneracy and estimate $L$ and $R_\mathrm{P}$ at least for the inner two  components. Given the possible contamination of the absorption line profiles by the outer Be star, we used the temperature estimates from the 4471/4481 line depth ratio and searched for combinations of $L$ and $R_\mathrm{P}$ that would result in the correct $H$-band flux ratio. In this way, we arrived at the best-fit parameters of $R_\mathrm{P, Aa} = 4.0$\,\Rsun, $L_\mathrm{Aa} = 600$\,\Lsun~and $R_\mathrm{P, Ab} = 2.5$\,\Rsun, $L_\mathrm{Ab} = 400$\,\Lsun. For these parameter values and the measured $v \sin{i}$, the rotation rates are $W_\mathrm{Aa} = 0.37$ and $W_\mathrm{Ab} = 0.56$, where $W$ is the ratio of the equatorial rotational velocity to the Keplerian velocity at the equator \citep[see Sect.~2.3.1 of][]{2013A&ARv..21...69R}. Again, we stress that these are only rough estimates and new data are needed to improve upon them.

\section{Discussion}
\label{sec:discussion}

\subsection{Dynamical and evolutionary state of the system and formation scenarios}

\begin{figure}[t]
   \centering
   \includegraphics[width=0.5\textwidth]{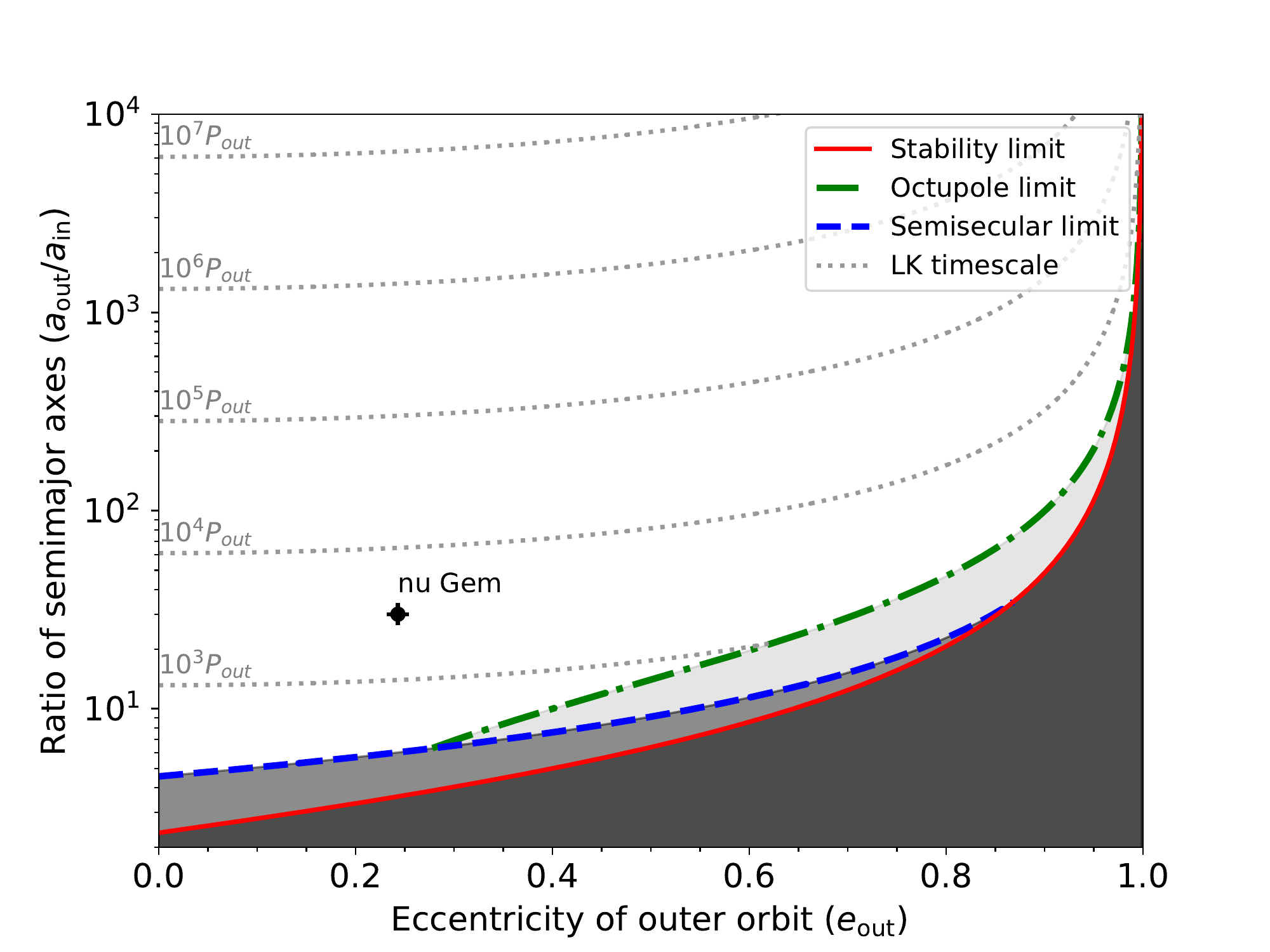}
   \caption{\label{fig:toonen_fig4} Dynamical parameter space of the $\nu$~Gem triple system. The illustration is the equivalent of Fig.~4a in \citet{2020A&A...640A..16T} and was created using their online tool at \url{https://bndr.it/wr64f}. The gray shaded area below the red line marks the dynamically unstable domain. Systems between the red and the dotted blue line are semisecularly stable (with maximum eccentricity of the inner pair assumed to be 0.99). Dotted lines trace Lidov-Kozai cycles of constant length (in units of the outer period). For plotting the octupole limit the octupole parameter $\epsilon_\mathrm{oct}$ was assumed to be 0.001 \citep[see Sect.~2.2.3 of ][]{2020A&A...640A..16T}.}
\end{figure}

The evolution of stellar triple systems is governed by the interplay of three-body dynamics, nuclear evolution of the stellar components, their rotation, mass loss, and mutual interactions. In addition to all the processes relevant to the evolution of single stars and binaries, there are aspects specific to triple stars, which most notably include the issue of dynamical stability, the (eccentric) Lidov-Kozai (LK) mechanism, and precession of the rotation axes.  A detailed analysis of the evolutionary state of the system is beyond the scope of this work but would probably require an extension of the observational time baselines by new spectroscopy and/or interferometry to make significant progress.  
In the combined orbital solutions for $\nu$~Gem (where solution C4 achieves the better agreement with the disentangled spectral line profiles) the ratio of outer to inner semimajor axis $\bar{a}'/\bar{a}$ is around 30. The stability criterion of \citet{2001MNRAS.321..398M} shows that, in its current configuration, $\nu$~Gem is well within the stable regime (Fig.~\ref{fig:toonen_fig4}). The semisecular regime in which the timescale of eccentricity oscillations of the inner system are comparable to the orbital timescales is also not relevant in this case as can be seen from the blue curve in Fig.~\ref{fig:toonen_fig4}. 

The LK mechanism in hierarchical triple systems acts to induce oscillations in the eccentricity of the inner pair and the relative inclination of the orbits by the dynamical influence from the distant third component \citep{1962P&SS....9..719L, 1962AJ.....67..591K, 2016ARA&A..54..441N}. This happens on timescales much longer than the outer orbital period. Fig.~\ref{fig:toonen_fig4} shows that $\nu$~Gem is in the standard (quadrupole) LK regime where the timescale of the LK oscillations is several thousand outer orbital periods, which is rather short compared to the evolutionary timescale of single stars.  Furthermore, the eccentric (octupole) LK mechanism (marked by the green curve in Fig.~\ref{fig:toonen_fig4}), which is capable of flipping the orbits from co-rotating to counter-rotating (and vice versa), does not seem to be of importance for $\nu$~Gem, although it could have been so in the past. However, wind-induced mass loss from the inner components should act to widen the inner orbit over time, which means that the inner orbit was probably more compact in the past (so that the $\bar{a}'/\bar{a}$ ratio was higher), thus making the inner orbit less prone to be influenced by the eccentric LK mechanism \citep{2016ARA&A..54..441N}. 

The low mutual inclination and the co-directional orbits of solution C4 are in agreement with this, and it is reasonable to assume that the system originated from fragmentation of its parental molecular cloud. In addition, the edge-on view of the outer Be star corroborates the notion of a rough alignment of spin and orbital angular momentum (see also next subsection), and the relatively strong rotational broadening of the spectral lines of the two inner stars shows that their spin misalignment, too, is not large.  Currently, the signs of the three spins are unknown but spectro-interferometry can determine that of the Be star by resolving its disk. By contrast, solution C2, which is difficult to reconcile with the disentangled line profiles and is also characterized by counter-rotating orbits, would favor the idea that the outer Be star was added to the $\nu$~Gem system by gravitational capture in the parental (now dispersed) star cluster.

The LK mechanism also induces precession of the argument of periastron of the inner orbit, although the timescale is too long to be noticeable over the timespan of our dataset (Fig.~\ref{fig:toonen_fig4}).  Other causes of apsidal precession include short-range forces from general relativistic effects, tidal distortions, and component rotation. The latter effects act in the opposite direction than the LK mechanism and are particularly important for very close and eccentric binaries \citep{2016ARA&A..54..441N}. In eccentric orbits, so-called heartbeats and orbitally-resonant stellar oscillations may also be excited \citep{2012ApJ...753...86T}. However, the Roche-lobe filling factor of the inner binary is only a few percent, rendering tidal influences in $\nu$~Gem presently likely negligible. The inclination of the total angular momentum of both orbits in solution C4 is 76.2{\degree}, and the plane of the inner orbit precesses about it with a semi-amplitude of 9.2{\degree} and a period of 8238 years \citep[according to][]{1975A&A....42..229S,2015MNRAS.449.1691B}.  In solution C2, the inclination of the total angular momentum is 72.9{\degree}, and the angular momentum precesses about it with an opening angle of 149.5{\degree} and a period of 11217 years. Even in this faster case, the present rate of change of the inclination of the inner orbit is 0.6{\degree} per hundred years so that it is not measurable in our current data.

\begin{figure}[t]
   \centering
   \includegraphics[width=0.5\textwidth]{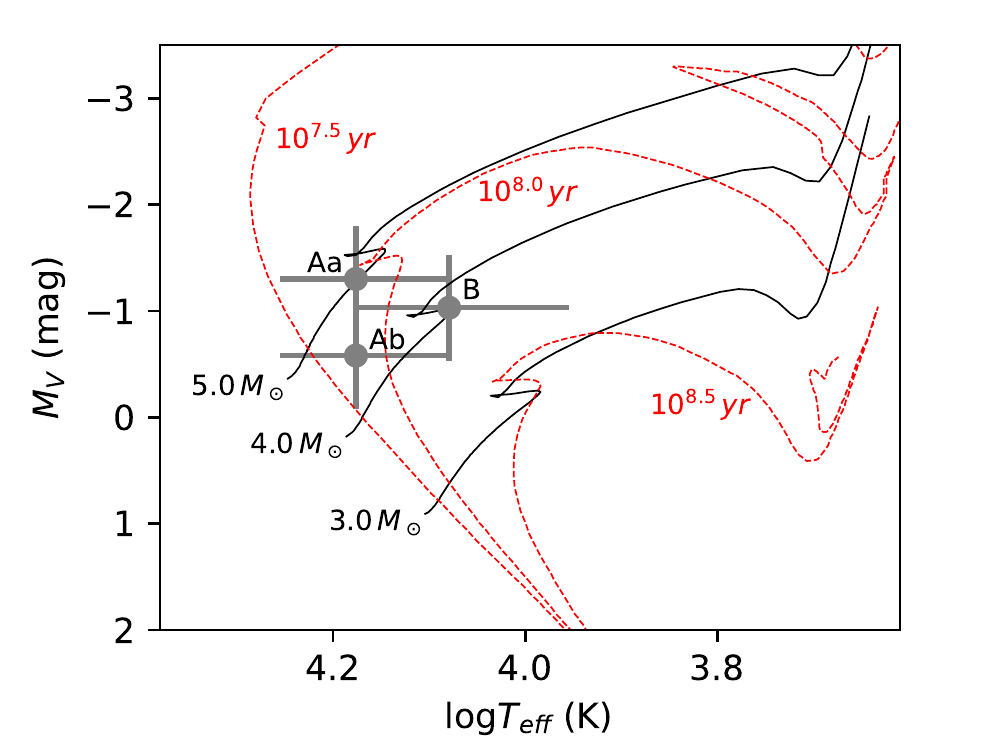}%
   \caption{\label{fig:HR_diagram} HR diagram showing evolutionary tracks for different masses (solid black) and isochrones for different ages (dotted red) from \citet{2012A&A...537A.146E}. The effective temperatures and absolute magnitudes were derived in Sect.~\ref{sec:specmodel} and here we adopt conservative error bars of 0.5 mag for the latter. The components appear to be roughly coeval and formed with similar ZAMS masses.
   }
\end{figure}

In order to assess the evolutionary state of $\nu$~Gem, we used the evolutionary tracks and isochrones for rotating stars at solar metallicity from \citet{2012A&A...537A.146E} for comparison with the derived parameters of the component stars from solution C4. The resulting HR diagram (Fig.~\ref{fig:HR_diagram}) is compatible with initial component masses of $\sim 4$--$5$\,{\Msun} and a common age of $\sim 10^{8.0}$\,yr. However, we note that the evolutionary models somewhat arbitrarily assume a zero-age main sequence (ZAMS) rotation rate of $v_\mathrm{initial} / v_\mathrm{critical} = 0.4$ (among other simplifying assumptions), while the individual components presently rotate at different rates despite having very similar masses.

\subsection{The outer Be star and its circumstellar disk}
\label{sec:discussion-Be}

Mass loss from Be stars is probably strongly equatorially enhanced, and the decretion disks formed by Be stars are co-planar with the central star's equator \citep{2013A&ARv..21...69R}. Without replenishment, Be disks are typically destroyed within about a year.  Therefore, the Be disk in $\nu$~Gem is an excellent tracer of the present-day inclination of the Be star.  Unfortunately, the available interferometric data do not constrain the ellipticity of the on-sky projection of the Be disk, but the inclination must be high because the formation of shell lines requires that the line of sight passes through the disk.  If it is assumed that spin and orbital momentum of the outer Be star are aligned, the inclination of the orbit of the Be star of ($75.9\pm0.1$)\,{\degree} implies a disk opening angle of at least $2 \times 14${\degree} (measured from midplane). This is at the high end of, but not in conflict with, the values typically found in Be-shell stars \citep{2006A&A...459..137R, 2014ApJ...795...82S}. 

Recent observations have suggested that most, if not all, Be stars could have close and hard-to-detect companions which are the remnants of past mass transfer that led to the spin-up of the Be star \citep[e.g.][]{2017A&A...601A..74K, 2018ApJ...853..156W, 2018A&A...615A..30S, Klement2019, 2021arXiv210313642W}.
\citet{2021AJ....161...40G} considered a low-mass low-luminosity companion to the Be star as one possible explanation of the disagreement between the inner Aa+Ab orbit determined from the inner astrometry and from the reflex wobble motion imprinted by the motions of these two stars on the orbit of the outer Be star.  However, they did not have spectra to further investigate the binary hypothesis for the Be star.  To this effect, we performed a Lomb-Scargle period analysis of the $O-C$ residuals of the outer RV curve (right panel of Fig.~\ref{fig:RV_curves}). No periodicity was detected so that it remains unknown whether component B actually is a binary Ba+Bb in which Ba would be the Be star and Bb a faint low-mass star.

The $\sim7.5$-year cycle length of the $V/R$ oscillations of H$\alpha$ (Sect.~\ref{sec:spec_variability}; Figs.~\ref{fig:Ha_linear} and \ref{fig:phased}) is not commensurate with both the inner and the outer orbital period in $\nu$~Gem. In principle, the putative companion of the Be star could excite orbitally resonant two-armed disk oscillations \citep[$m=2$ density waves;][]{2018MNRAS.473.3039P} that would reveal themselves through $V/R$ variability in double-peaked emission lines.  Their cycle length would be expected to be around half the orbital period of the putative companion.  If the latter has, in fact, dumped mass on the Be star, its orbital period should be less than about a year.  Therefore, the $V/R$ variability is probably not an indirect indicator of the presence of a companion of the Be star. Irrespective of the origin and nature of the $V/R$ variability, the underlying structural variations of the disk could appear in the interferometric measurements as variable resolved features.  As already discussed by \citet{2021AJ....161...40G}, this would be an alternative explanation of the residuals of their solution and obviate the need for invoking the presence of a fourth component in the system.

The system architecture of $\nu$~Gem is the same as that inferred by \citet{2020A&A...637L...3R} for HR\,6819 except that in their model for HR\,6819 (i) the inner binary does not consist of two luminous stars but one star plus a black hole (BH) and (ii) the orbital period of the outer Be star is undetermined but of the order of decades.  \citet{2020ApJ...898L..44G} found $V/R$ variations in the disk of the Be star with what \citeauthor{2020A&A...637L...3R} had identified as the inner orbital period.  Therefore, \citet{2020ApJ...898L..44G} attributed the origin of the $V/R$ variations to the second star, which they placed in a short-period orbit about the Be star, and concluded that the system is a binary without a BH.  While this interpretation is not disproved,  Fig.\,\ref{fig:phased} (panel 2) proves that it is not necessary:  The orbital motions in the inner binary can induce $V/R$ variations in the disk of the outer Be star, either as a physical density wave or just apparently as the result of the shifting of the underlying absorption lines.  Another proposed triple system with a BH and an outer Be star is LB-1 \citep{2020A&A...637L...3R} for which, too, a binary model based on $V/R$ and associated RV variations has been presented \citep{2020ApJ...900...42L}. 

As mentioned in the Introduction, there is a strong lack of MS companions to Be stars in general and of eclipsing Be stars in particular.  Therefore, to the best of our knowledge, the mass determined for component B in $\nu$~Gem is only the second ever dynamical measurement of the mass of a classical Be star.  \citep[The first case is $\varphi$~Per the companion to which, however, is a hot subdwarf,][] {2015A&A...577A..51M}

\section{Summary and conclusions}
\label{sec:conclusions}

$\nu$~Gem is a triple system consisting of three MS stars with nearly equal masses of 3.3\,M$_\odot$; one of them is a classical Be star.  Two of the stars form an inner binary with orbital period 53.8\,d and the third one is the Be star and orbits this pair with a period of 19.1 years.  At 0.056, the eccentricity of the inner binary is small whereas the outer orbit is more significantly non-circular with $e = 0.24 \pm 0.08$.  The relative inclination between the two orbits is 10 degrees, and the plane of the circumstellar disk of the Be star is aligned with the orbit to a similar level.  

In the inner binary, interferometry found both a brighter and a fainter star, and spectroscopy detected a broad- as well as a narrower-lined star.  The fainter star has the broader lines. The difference by a factor of two in brightness between the two inner stars, which have the same age, mass, and supposedly composition, is not finally explained but may involve their fairly different rotation rates. This requires more specific modeling.  At the fractional critical rotation rate of 0.56 determined in Sect.\,\ref{sec:specmodel} for the fainter star, gravity darkening of this magnitude is not expected, and there is little room for a much higher inclination unless a strong spin-orbit misalignment is postulated.

The above describes our preferred solution C4 (Table\,\ref{tab3}). 
However, the 180\degree\ degeneracy in $\omega$ of interferometric data makes the identities of the bright and the faint star interchangeable, and solution C2 (Table\,\ref{tab3}) in which the two orbits are counter-rotating and the masses are slightly different fits the orbital data equally well.  In this latter case, our modeling of the disentangled spectra leads to unrealistic physical properties so that we prefer solution C4.

This paper has laid the foundation to much future work:  It will be attractive to continue the interferometric monitoring as interferometry seems to be more sensitive to the presence of a putative companion to the Be star.  A firm conclusion concerning its presence or absence would be an important datum in understanding the diversity of processes that lead to the formation of Be stars.  The spectro-interferometric determination of the direction of rotation of the disk around the Be star may link the formation of this Be star to that of the $\nu$~Gem triple system as a whole.  At the given spectral types of all three component stars, only very few spectral lines other than the Balmer series appear in the optical.  Very high-quality spectra covering the orbital period of the inner binary will substantially widen the scope of spectral disentangling in order to search for spectral signatures of past interactions. Extreme adaptive optics may help to separate the spectrum of the Be star from those of the inner binary. An important objective for amateur spectroscopy is the comparison to the Be-star's orbital period of the cyclicity of the $V/R$ activity in the circumstellar disk.  

$\nu$~Gem is presently stable, and there are no hints of a tumultuous past.  This may mean that the system still traces the angular-momentum distribution on different spatial scales in the molecular cloud from which the system formed. In solution C4, the two orbits are co-directional so that there is no indication that the outer Be star was captured.  By contrast, capture may have to be considered for solution C2 in which the two orbits are counter-rotating. Another important detail to be clarified is again whether the present high angular momentum of the Be star is primordial or was acquired in a close binary interaction.  The fact that all three known components have nearly identical masses may be another motivation for studying possible genesis models.  


\acknowledgments{
This work is based upon observations obtained with the Georgia State University Center for High Angular Resolution Astronomy Array at Mount Wilson Observatory.  The CHARA Array is supported by the National Science Foundation under Grant No. AST-1636624 and AST-1715788.  Institutional support has been provided from the GSU College of Arts and Sciences and the GSU Office of the Vice President for Research and Economic Development.
MIRC-X received funding from the European Research Council (ERC) under the European Union's Horizon 2020 research and innovation programme (Grant No.\ 639889) and N.A., C.L.D., and S.K.\ acknowledge funding from the same grant.
A.L. received support from an STFC studentship (No.\ 630008203).
R.K. is grateful for a postdoctoral associateship funded by the Provost’s Office of Georgia State University. The research of R.K. is also supported by the National Science Foundation under Grant No. AST-1908026.
M.H. is supported by an ESO fellowship.
This research is based on spectra from the Ond{\v r}ejov 2m Perek Telescope provided by the services of the Czech Virtual Observatory; the observers were  Z.\ Bardon, J.\ Fuchs, P.\ Hadrava, J.\ Havelka, A.\ Kawka, D.\ Kor\v{c}\'{a}kov\'{a}, L.\ Kotkov\'{a}-\v{S}arounov\'{a}, P.\ Koubsk\'{y}, P.\ N\'{e}meth, M.\  Netolick\'{y}, J.\ Polster, L.\ \v{R}ezba, P.\ \v{S}koda, M.\ \v{S}lechta, J.\ Sloup, M.\ Tlamicha, S.\ \v{S}tefl, F.\ Votruba, and F.\ \v{Z}d'\'{a}rsk\'{y}.  
This work has made use of the BeSS database, operated at LESIA, Observatoire de Meudon, France: http://basebe.obspm.fr, of NASA's Astrophysics Database \citep{2000A&AS..143...41K}, and of the  SIMBAD database, operated at CDS, Strasbourg, France \citep{2000A&AS..143....9W}.  The authors sincerely thank the BeSS observers E.\ Barbotin, E.\ Bertrand, M.\ Bonnement, E.\ Bryssinck, C.\ Buil, F.\ Cochard, V.\ Desnoux, A.\ Favaro, P.\ Fosanelli, O.\ Garde, T.\ Garrel, K.\ Graham, J.\ Guarro Fl\'o, A.\ Heidemann, F.\ Houpert, J.\ Jacquinot, T.\ Lemoult, M.\  Leonardi, G.\ Martineau, B.\ Mauclaire, C.\ Sawicki, O.\ Thizy, and S.\ Ubaud for their dedication and efforts.  This research has made use of NASA’s Astrophysics Data System. 
Based on INES data from the IUE satellite.
}

\facilities{CHARA, OO:2, IUE}

\software{Astropy \citep{astropy:2013, astropy:2018}, Matplotlib \citep{Hunter:2007}, PyAstronomy\footnote{\url{https://github.com/sczesla/PyAstronomy)}} \citep{pya}, RadVel: General toolkit for modeling Radial Velocities\footnote{\url{https://zenodo.org/record/580821}} \citep{2018PASP..130d4504F}}

\bibliography{nu_gem.bib}

\appendix

\section{Archival speckle measurements}

The archival speckle measurements used in this study are summarized in Table~\ref{tab:speckle}.

\begin{deluxetable}{LCC}[h!]
\tablenum{A.1}
\tablecaption{Results from speckle measurements retrieved from the WDS catalog showing the separation and PA of component B relative to the photocenter of Aa+Ab. \label{tab:speckle}}
\tablewidth{0pt}
\tablehead{
\colhead{MJD} & \colhead{separation} & \colhead{PA} \\
\nocolhead{MJD} & \colhead{[mas]} & \colhead{[\degree]}
}
\startdata
43091.157 & 58.0 & 109.1 \\
43092.179 & 58.0 & 109.5 \\
43174.943 & 64.0 & 111.9 \\
43561.881 & 83.0 & 116.3 \\
43941.842 & 101.0 & 126.1 \\
44155.217 & 95.0 & 122.2 \\
44294.958 & 92.0 & 123.8 \\
44504.206 & 99.0 & 124.2 \\
44560.197 & 92.0 & 124.6 \\
45245.173 & 81.0 & 137.0 \\
45351.933 & 74.0 & 141.8 \\
45720.974 & 51.0 & 141.3 \\
45722.982 & 57.0 & 145.7 \\
46371.141 & 36.0 & 183.2 \\
47609.787 & 56.0 & 288.2 \\
48167.147 & 64.0 & 298.3 \\
48584.034 & 53.0 & 309.8 \\
49670.118 & 45.0 & 91.0 \\
50497.757 & 84.0 & 117.3 \\
54747.862 & 60.5 & 290.8 \\
54926.428 & 60.9 & 293.7 \\
55574.587 & 55.8 & 310.6 \\
56678.605 & 34.6 & 93.8 \\
56933.909 & 55.3 & 106.4 \\
56933.909 & 54.2 & 108.1 \\
57060.502 & 66.1 & 105.7 \\
57060.502 & 64.0 & 109.0 \\
57354.632 & 78.7 & 113.8 \\
57354.632 & 78.7 & 114.4 \\
57738.976 & 91.6 & 117.7 \\
\enddata
\end{deluxetable}

\section{Archival polarimetric measurements}

\setcounter{figure}{0} \renewcommand{\thefigure}{B.\arabic{figure}}

\begin{figure}[t]
   \centering \includegraphics[width=0.5\textwidth]{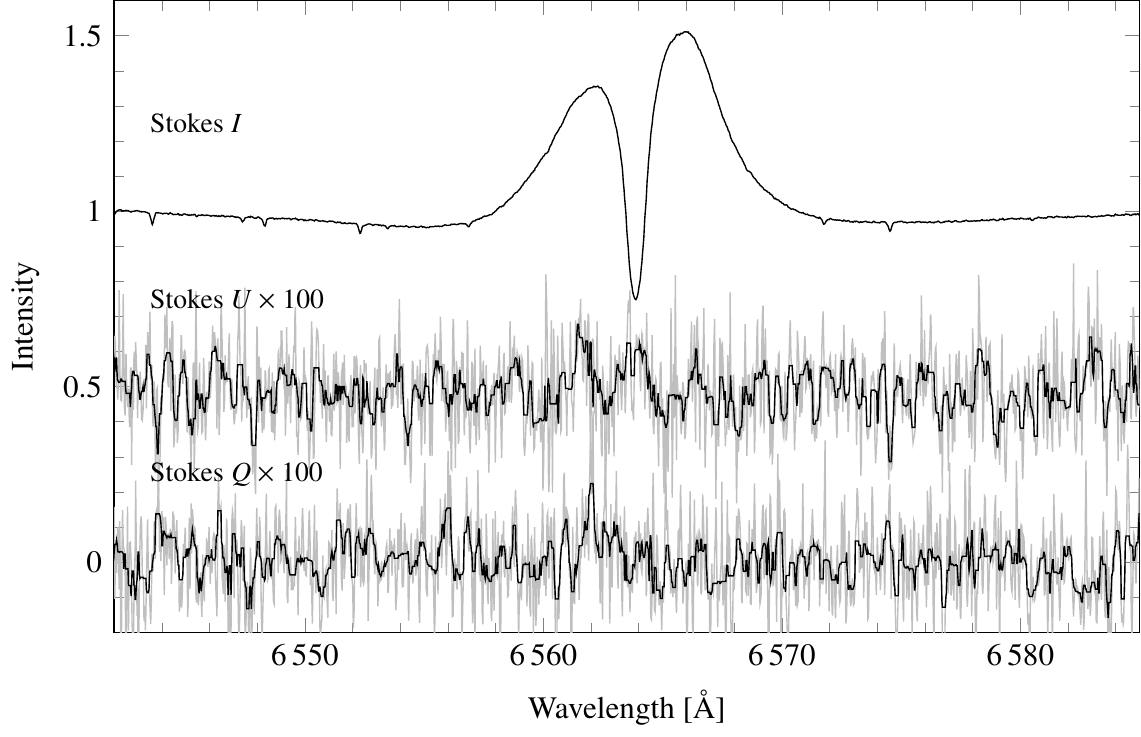}
   \caption{\label{fig:ESPpol}ESPaDOnS polarimetric data across H$\alpha$. For $Q$ and $U$ the original data are plotted in light gray and in black after median filtering over 5 pixels.  Both multiplied by a factor of 100, and $U$ is shifted upwards by 0.5.}
\end{figure}

The archival polarimetric measurements used in this study are summarized in Table~\ref{tab:hpol} and Fig.~\ref{fig:ESPpol}.

\begin{deluxetable}{LCCCCC}[h!]
\tablenum{B.1}
\tablecaption{HPOL measurements taken from {\tt http://www.sal.wisc.edu/HPOL/tgts/Nu-Gem.html}\label{tab:hpol}}
\tablewidth{0pt}
\tablehead{
\colhead{Band} & \colhead{$Q$} & \colhead{$U$} &  \colhead{Error} & \colhead{Pol.\ degree} & \colhead{PA} \\
\nocolhead{Band} & \colhead{[\%]} &  \colhead{[\%]} & \colhead{[\%]} & \colhead{[\%]} &\colhead{[\degree]}
}
\startdata
\multicolumn{6}{c}{MJD=48683.19 (Reticon)}\\
$Ux$ &   0.1917 &   0.0534 &   0.0135 &   0.1990 & 7.786  \\  
$B$  &   0.2166 &   0.1477 &   0.0033 &   0.2621 & 17.148  \\  
$V$  &   0.2004 &   0.1499 &   0.0032 &   0.2502 & 18.399  \\  
$R$  &   0.1951 &   0.1640 &   0.0047 &   0.2549 & 20.026  \\  
$I$  &   0.1936 &   0.0717 &   0.0199 &   0.2065 & 10.164  \\  
\hline \noalign{\smallskip}
\multicolumn{6}{c}{MJD=48712.05 (Reticon)}\\                        
$Ux$ &   0.2108 &   0.0978 &   0.0071 &   0.2324 & 12.440  \\  
$B$  &   0.2039 &   0.1483 &   0.0020 &   0.2521 & 18.017  \\  
$V$  &   0.2082 &   0.1564 &   0.0021 &   0.2604 & 18.460  \\  
$R$  &   0.2086 &   0.1529 &   0.0028 &   0.2586 & 18.119  \\  
$I$  &   0.1650 &   0.2318 &   0.0127 &   0.2846 & 27.277  \\  
\hline \noalign{\smallskip}
\multicolumn{6}{c}{MJD=51159.85 \& 51159.80 (CCD Blue \& Red)}\\
$Ux$ &   0.2073 &   0.1060 &   0.0152 &   0.2329 & 13.543  \\  
$B$  &   0.2537 &   0.1451 &   0.0044 &   0.2922 & 14.883  \\  
$V$  &   0.2429 &   0.1311 &   0.0028 &   0.2761 & 14.178  \\  
$R$  &   0.2261 &   0.1417 &   0.0022 &   0.2668 & 16.039  \\  
$I$  &   0.2168 &   0.1523 &   0.0024 &   0.2649 & 17.539  \\  
\hline \noalign{\smallskip}
\multicolumn{6}{c}{MJD=51195.75  \& 51195.70 (CCD Blue \& Red)} \\
$Ux$ &   0.1599 &   0.1063 &   0.0123 &   0.1920 & 16.802  \\  
$B$  &   0.2164 &   0.1246 &   0.0039 &   0.2498 & 14.967  \\  
$V$  &   0.2174 &   0.1359 &   0.0024 &   0.2564 & 16.004  \\  
$R$  &   0.2240 &   0.1365 &   0.0014 &   0.2623 & 15.679  \\  
$I$  &   0.2084 &   0.1576 &   0.0016 &   0.2613 & 18.550  \\  
\hline \noalign{\smallskip}
\multicolumn{6}{c}{MJD=51957.80  \& 51957.76  (CCD Blue \& Red)  }\\
$Ux$ &   0.1439 &   0.6932 &   0.1805 &   0.7080 & 39.136  \\  
$B$  &   0.1786 &   0.1858 &   0.0118 &   0.2577 & 23.068  \\  
$V$  &   0.2256 &   0.1283 &   0.0047 &   0.2595 & 14.807  \\  
$R$  &   0.2140 &   0.1297 &   0.0026 &   0.2502 & 15.616  \\  
$I$  &   0.1859 &   0.1470 &   0.0023 &   0.2370 & 19.172  \\  
\hline \noalign{\smallskip}
\enddata 
\end{deluxetable}

\begin{deluxetable}{LCCCCC}[h!]
\tablenum{B.1 cont.}
\tablecaption{HPOL measurements taken from {\tt http://www.sal.wisc.edu/HPOL/tgts/Nu-Gem.html}}
\tablewidth{0pt}
\tablehead{
\colhead{Band} & \colhead{$Q$} & \colhead{$U$} &  \colhead{Error} & \colhead{Pol.\ degree} & \colhead{PA} \\
\nocolhead{Band} & \colhead{[\%]} &  \colhead{[\%]} & \colhead{[\%]} & \colhead{[\%]} &\colhead{[\degree]}
}
\startdata
\multicolumn{6}{c}{MJD=52263.98  \& 52263.95  (CCD Blue \& Red)}\\
$Ux$ &   -0.0462 &  0.2949   & 0.2075 &   0.2985 & 49.453  \\  
$B$  &   0.2813  &  0.1520   & 0.0116 &   0.3197 & 14.190  \\  
$V$  &   0.2038  &  0.1763   & 0.0052 &   0.2695 & 20.432  \\  
$R$  &   0.2191  &  0.1896   & 0.0033 &   0.2898 & 20.433  \\  
$I$  &   0.2063  &  0.1590   & 0.0032 &   0.2605 & 18.813  \\  
\hline \noalign{\smallskip}
\multicolumn{6}{c}{MJD=52299.80 \& 52299.76    (CCD Blue \& Red)}\\
$Ux$ &   0.1333 &   0.0804 &   0.0233 &   0.1557 & 15.550  \\  
$B$  &   0.2078 &   0.1526 &   0.0053 &   0.2578 & 18.143  \\  
$V$  &   0.2397 &   0.1343 &   0.0031 &   0.2747 & 14.630  \\  
$R$  &   0.2303 &   0.1483 &   0.0020 &   0.2739 & 16.387  \\  
$I$  &   0.2147 &   0.1453 &   0.0020 &   0.2592 & 17.045  \\  
\hline \noalign{\smallskip}
\multicolumn{6}{c}{MJD=52319.79 \& 52319.73    (CCD Blue \& Red)}\\
$Ux$ &   0.1209 &   -0.0148 &  0.0462 &   0.1218 & 176.520 \\   
$B$  &   0.2455 &   0.1349  &  0.0065 &   0.2801 & 14.388  \\  
$V$  &   0.2363 &   0.1517  &  0.0034 &   0.2808 & 16.348  \\  
$R$  &   0.2281 &   0.1710  &  0.0020 &   0.2851 & 18.433  \\  
$I$  &   0.2054 &   0.1615  &  0.0020 &   0.2613 & 19.086  \\  
\hline \noalign{\smallskip}
\multicolumn{6}{c}{MJD=53049.64 \& 53049.61    (CCD Blue \& Red)}\\
$Ux$ &   0.2275 &   0.1354 &   0.0161 &   0.2647 & 15.377  \\  
$B$  &   0.2292 &   0.1292 &   0.0057 &   0.2631 & 14.702  \\  
$V$  &   0.2356 &   0.1481 &   0.0036 &   0.2783 & 16.074  \\  
$R$  &   0.2378 &   0.1526 &   0.0024 &   0.2826 & 16.346  \\  
$I$  &   0.2239 &   0.1318 &   0.0025 &   0.2598 & 15.245  \\  
\enddata
\end{deluxetable}


\end{document}